\documentclass[twocolumn]{aastex63}

\usepackage{amssymb}
\usepackage{tabularx}
\usepackage{rotating}
\usepackage{xcolor}

\usepackage{amsmath}

\usepackage{lineno}

\received{July 19, 2019}
\revised{Oct. 31, 2019}
\accepted{Nov. 7, 2019}

\shorttitle{Saturn's internal magnetic field from Cassini Grand Finale}
\shortauthors{Cao et al.}

\begin{document}

\title{The landscape of Saturn's internal magnetic field from the Cassini Grand Finale}

\correspondingauthor{Hao Cao}
\email{haocao@fas.harvard.edu}

\author[0000-0002-6917-8363]{Hao Cao}
\affiliation{Department of Earth and Planetary Sciences, Harvard University, 20 Oxford Street, Cambridge, MA 02138, USA}

\affiliation{Division of Geological and Planetary Sciences, California Institute of Technology, Pasadena, CA 91125, USA}

\affiliation{Physics Department, The Blackett Laboratory, Imperial College London, London, SW7 2AZ, UK}

\author[0000-0002-9658-8085]{Michele K. Dougherty}
\affiliation{Physics Department, The Blackett Laboratory, Imperial College London, London, SW7 2AZ, UK}

\author[0000-0002-9154-7081]{Gregory J. Hunt}
\affiliation{Physics Department, The Blackett Laboratory, Imperial College London, London, SW7 2AZ, UK}

\author[0000-0001-7442-4154]{Gabrielle Provan}
\affiliation{Department of Physics and Astronomy, University of Leicester, Leicester, LE1 7RH, UK}

\author[0000-0002-4041-0034]{Stanley W.H. Cowley}
\affiliation{Department of Physics and Astronomy, University of Leicester, Leicester, LE1 7RH, UK}

\author[0000-0002-9456-0345]{Emma J. Bunce}
\affiliation{Department of Physics and Astronomy, University of Leicester, Leicester, LE1 7RH, UK}

\author{Stephen Kellock}  
\affiliation{Physics Department, The Blackett Laboratory, Imperial College London, London, SW7 2AZ, UK}

\author[0000-0001-9432-7159]{David J. Stevenson}
\affiliation{Division of Geological and Planetary Sciences, California Institute of Technology, Pasadena, CA 91125, USA}

\begin{abstract}
The Cassini mission entered the Grand Finale phase in April 2017 and executed 22.5 highly inclined, close-in orbits around Saturn before diving into the planet on September 15th 2017. Here we present our analysis of the Cassini Grand Finale magnetometer (MAG) dataset, focusing on Saturn's internal magnetic field. These measurements demonstrate that Saturn's internal magnetic field is exceptionally axisymmetric, with a dipole tilt less than 0.007 degrees {(25.2 arcsecs)}. Saturn's magnetic equator was directly measured to be shifted northward by $\sim$ 0.0468 $\pm$ 0.00043 (1$\sigma$) $R_S$, 2820 $\pm$ 26 $km$, at cylindrical radial distances between 1.034 and 1.069 $R_S$ from the spin-axis. Although almost perfectly axisymmetric, Saturn's internal magnetic field exhibits features on many characteristic length scales in the latitudinal direction. Examining $B_r$ at {the $a=0.75$ $R_S$, $c=0.6993$ $R_S$ isobaric surface}, the degree 4 to 11 contributions correspond to latitudinally banded {magnetic perturbations} with characteristic width $\sim$ 15$^\circ$, similar to that of the off-equatorial zonal jets observed in the atmosphere of Saturn. Saturn's internal magnetic field beyond 60$^\circ$, in particular the small-scale features, are less well constrained by the available measurements, mainly due to incomplete spatial coverage in the polar region. Magnetic fields associated with the ionospheric Hall currents were estimated and found to contribute less than 2.5 $nT$ to Gauss coefficients beyond degree 3. The magneto-disk field features orbit-to-orbit variations between {12 $nT$ and 15.4 $nT$} along the close-in part of Grand Finale orbits, offering an opportunity to measure the electromagnetic induction response from the interior of Saturn. {A stably stratified layer thicker than 2500 $km$ likely exists above Saturn's deep dynamo to filter out the non-axisymmetric internal magnetic field. A heat transport mechanism other than pure conduction, e.g. double diffusive convection, must be operating within this layer to be compatible with Saturn's observed luminosity. The latitudinally banded magnetic perturbations likely arise from a shallow secondary dynamo action with latitudinally banded differential rotation in the semi-conducting layer.}
\end{abstract}


\section{Introduction}
\label{intro}

Intrinsic magnetic field is a fundamental property of a planet. Not only is it a key factor in determining the electromagnetic environment of a planetary body, it also serves as a key diagnostic of the interior structure and dynamics of the host planet \citep[]{Stevenson2003, Stevenson2010}. A strong planetary-scale magnetic field most likely originates from dynamo action within the planet, the operation of which requires a large volume of electrically conducting fluid and ``fast" and complex fluid motions \citep[]{SKR1966, SK1966, Parker1955, KR1980, RS1971, RK2013}. For gas giant dynamos, metallic hydrogen is the electrically conducting fluid, secular cooling drives ``fast" fluid motions, while the rapid background rotation promotes the generation of large-scale magnetic fields \citep[]{Christensen2010}. The warm interior conditions of the present-day Jupiter and Saturn makes the transition from molecular to metallic hydrogen a gradual process: the electrical conductivity rises rapidly yet continuously from negligible values in the 1-bar atmosphere to significant values in the Mbar region \citep[]{Weir1996, Liu2008}. The transition from magnetohydrodynamics (MHD) in the deep dynamo to hydrodynamics in the outer layers inside gas giants is also likely to be gradual \citep[]{CS2017Icarus}. It is generally believed that the transition from hydrodynamics to MHD underlies the transition from 100 $m/s$ rapid zonal flows in the non-conducting outer layer to $cm/s - mm/s$ slow deep dynamo flows inside the gas giants \citep[]{Kaspi18, Guillot18}. However, the physical mechanism of this dynamical transition, in particular that at mid-to-high latitudes, remains unknown. On the other hand, although fluid motions in the semi-conducting layer may not be able to sustain dynamo action on their own, they could modify the deep dynamo generated magnetic field and produce observable features outside the planet such as magnetic perturbations spatially correlated with deep zonal flows {\citep[]{Gastine2014,CS2017Icarus}} and time variation of the magnetic field (secular variation) \citep[]{Moore2019}. 

Saturn's magnetic field has been measured in-situ by four space missions, Pioneer 11 \citep[]{Smith1980, Acuna1980}, Voyager 1 \citep[]{Ness1981}, Voyager 2 \citep[]{Ness1982, Connerney1982}, and Cassini \citep[]{Dougherty2005, Burton2009, Cao2011,Cao2012, DC2018}. These measurements revealed an almost perfectly axisymmetric, dipole dominant internal magnetic field with non-negligible north-south asymmetry \citep[]{DC2018} and a highly dynamic magnetosphere filled with periodic phenomena whose frequencies are close to the rotational frequency of Saturn \citep[]{Andrews2012, Provan2018, Provan2019}. The periodic magnetic perturbations in Saturn's magnetosphere are referred to as Planetary Period Oscillations (PPOs). The search for departures from perfect axisymmetry in the internal magnetic field of Saturn is of great interest, since it could yield the true rotation period of the deep interior {\citep[see current values derived from different measurements and methods:][]{Anderson2007, Read2009, Mankovich2019, Militzer2019}} and provide key constraints on the dynamo process inside Saturn. However, this search is complicated by the existence of ionospheric and field-aligned currents (FACs) at Saturn, which feature both PPO and non-PPO components \citep[]{Hunt2014, Hunt2015, Hunt2018}. Here we would like to stress that the deep dynamo layer of Saturn rotates very much like that of a solid body from the view of observers in an inertial frame, since the expected $cm/s - mm/s$ differential rotation is only about one part in a million when compared to the $\sim$ 10 $km/s$ background rotation. 

Among the existing measurements, those from the Grand Finale phase of the Cassini mission (Table 1, Figs. 1 - 4) are the most sensitive to the internal magnetic field due to their proximity to the planet and the highly inclined orbit. So far, the analysis of Saturn's internal magnetic field has been mostly restricted to the traditional Gauss coefficients representation, in which the internal planetary magnetic field is expressed as a function of the Gauss coefficients $(g_n^m,h_n^m)$ with
\begin{equation}
B_{r,\theta,\phi}(r,\theta,\phi)=\sum_n\sum_m [g_n^m f^g_{r,\theta,\phi}(r,\theta,\phi) +h_n^m f^h_{r,\theta,\phi}(r,\theta,\phi)]
\label{eqn:B_gh}
\end{equation} 
where the functional form of $f^{g,h}_{r,\theta,\phi}$ can be easily found \citep[e.g., Eqns. 3 - 5 in][]{DC2018} and reproduced in Appendix A for convenience. An equivalent and likely more fundamental representation of the internal magnetic field of a planet is the Green's function which maps the internal magnetic field from the dynamo surface (or the planetary surface) to the field outside \citep[e.g.][]{GR1983, Backus1996, JC1997}:
\begin{equation}
B_{r,\theta,\phi}^{obs}(r,\theta,\phi)=\int_0^{2\pi}\int_0^{\pi} B^{r_D}_r(\theta',\phi')G_{r,\theta,\phi}(\mu)\sin\theta'd\theta'd\phi',
\label{eqn:B_Green}
\end{equation}
here $B_r^{r_D}$ is the radial component of the magnetic field at the dynamo surface (a spherical surface with $r=r_D$ {in the traditional geophysical formulation, see next paragraph for non-sphericity of isobaric surface inside Saturn}),  $B^{obs}_{r,\theta,\phi}$ are three components of the magnetic field measured above the {``dynamo surface", and $\mu$ is the cosine of the angle between the position vectors $\hat{r}$ and $\hat{r}'$ (see Appendix B for more details)}. The Green's function not only yields an equivalent description of the internal magnetic field, it also admits a simple and straightforward physical interpretation: it describes how sensitive the magnetic field measured outside the planet is to the field at different locations on the dynamo surface. The Green's function has been applied to analyzing the magnetic field of the Earth \citep[e.g.][]{JC1997,Jackson2007}, Mars \citep[]{Purucker2000, Moore2017JGR}, and Jupiter \citep[]{Moore2017GRL}. 

{ Saturn is the most oblate planet in the solar system, with a measured flattening $f=(a-c)/a = 9.8\%$ at the 1-bar surface, where $a$ and $c$ are the equatorial radius and polar radius respectively. The flattening of the interior isobaric surface decreases as the pressure level increases \citep[e.g., see Fig. 2 in][]{CS2017JGR}. According to the latest Saturn interior model \citep[]{Militzer2019} constrained by the Cassini Grand Finale gravity measurements \citep[]{Iess2019}, the flattening of the isobaric surface decreases to 6.76\% at $a=0.75 R_S$ and 5.88\% at $a= 0.65 R_S$. The isobaric surfaces of giant planets are not perfect ellipsoids due to their non-uniform density. The fractional deviation from ellipsoids $\Delta r/r$, however, are on the order of $10^{-3}$ or less for Jupiter and Saturn \citep[e.g., see Fig. 2 in][]{CS2017JGR, Militzer2019}, two orders of magnitude smaller than the dominant elliptical flattening. Thus, we treat the ``dynamo surface" as ellipsoids when evaluating the properties of Saturn's internal magnetic field.}

Here we report our analysis of the Cassini Grand Finale MAG dataset, focusing on Saturn's internal magnetic field. {It should be noted that the solution of Saturn's internal magnetic field were obtained with spherical basis function, such as the spherical harmonics and the Green's functions on a sphere. However, the non-spherical shape of Saturn's ``dynamo surface" was explicitly considered when evaluating the properties of the resultant internal magnetic field.} We have extended the analysis presented in \citet[]{DC2018} in several ways: i) MAG data from the last 12.5 Cassini Grand Finale orbits are analyzed here together with those presented in \citet[]{DC2018}, ii) an explicit search for internal non-axisymmetry is carried out, iii) the effect of incomplete spatial coverage is demonstrated with regularized inversion, and iv) Green's functions were employed in addition to the traditional Gauss coefficients in constructing models of Saturn's internal magnetic field, v) ionospheric current and their associated magnetic field are evaluated with a simple axisymmetric model, and vi) search for electromagnetic induction from the interior of Saturn and orbit-to-orbit varying ``internal" field is carried out. In section 2, we present the main characteristics of the trajectory of Cassini Grand Finale orbits and the MAG measurements. In section 3, we present the directly measured position of Saturn's magnetic equator and its spatial variations. In section 4, we present the sensitivity of Cassini Grand Finale MAG measurements to Saturn's axisymmetric internal magnetic field at depth. In section 5, we present inversion of Saturn's axisymmetric internal magnetic field with different methods. In section 6, we present a search for electromagnetic induction from the interior of Saturn. In section 7, we present the orbit-to-orbit variations in Saturn's ``internal" quadrupole magnetic moments. In section 8, we present a search for internal non-axisymmetry in Saturn's magnetic field. In section 9, we discuss the constraints and implications on Saturn's interior structure and dynamics. Section 10 presents a summary and outlook.

\section{Cassini Grand Finale trajectory and MAG measurements} %

\begin{figure}[h]
\includegraphics[width=0.475\textwidth]{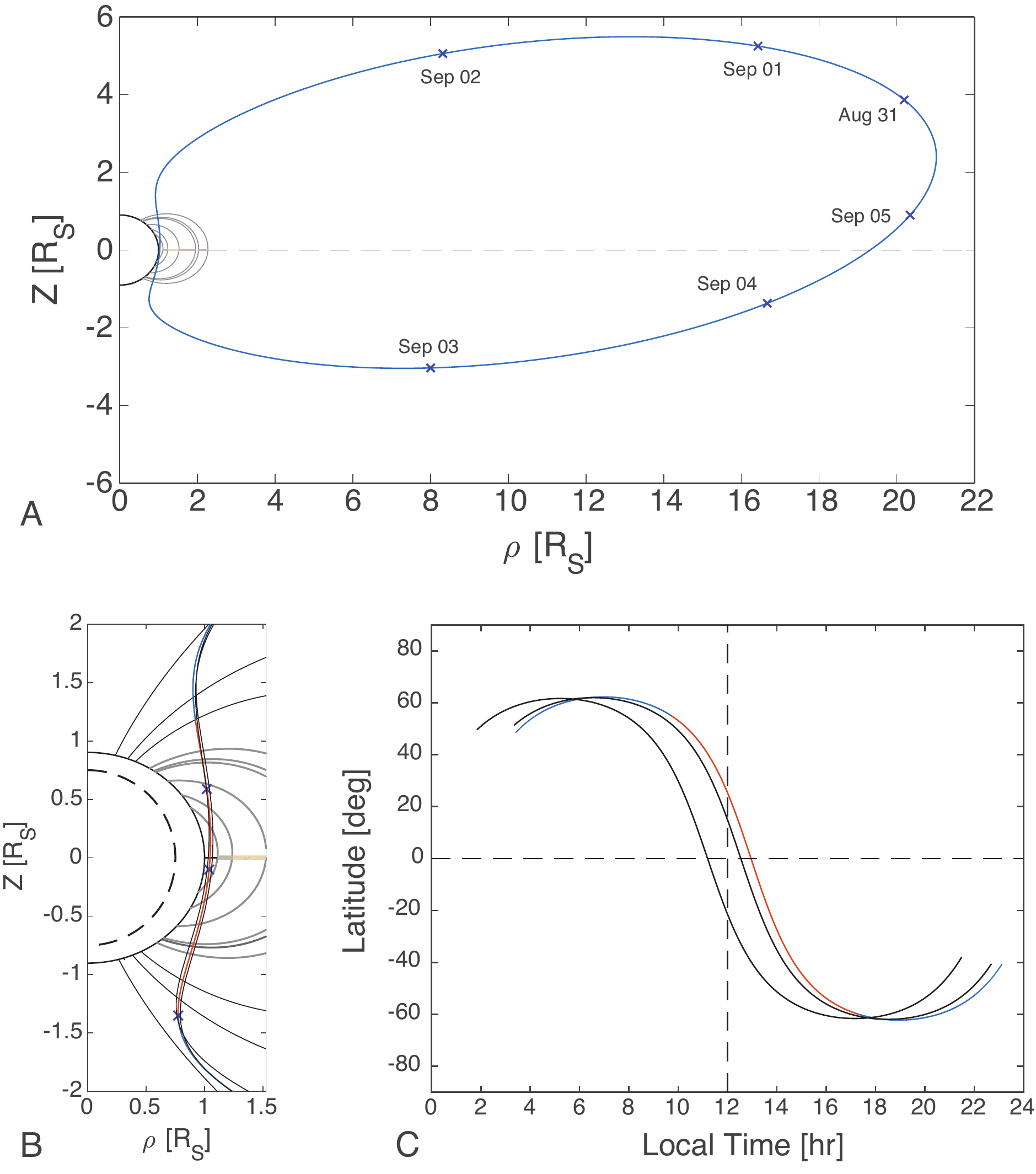}
\caption{Trajectory of typical Cassini Grand Finale orbits. In panel A, the trajectory of Rev 291 from apoapsis to apoapsis is projected onto the meridional plane in which $Z$ is along the spin-axis direction and $\rho$ is in the cylindrical radial direction. Panel B shows the close-in part of the trajectory from three Cassini Grand Finale orbits in the same projection. For the blue-red color-coded trajectory, the red part is when the measured magnetic field strength $>$ 10,000 $nT$. The dashed line shows $r$ = 0.75 $R_S$. Panels C shows the trajectory in latitude local time projection.}
\label{fig1}
\end{figure}

\begin{figure}[h]
\centering
%
\includegraphics[width=0.475\textwidth]{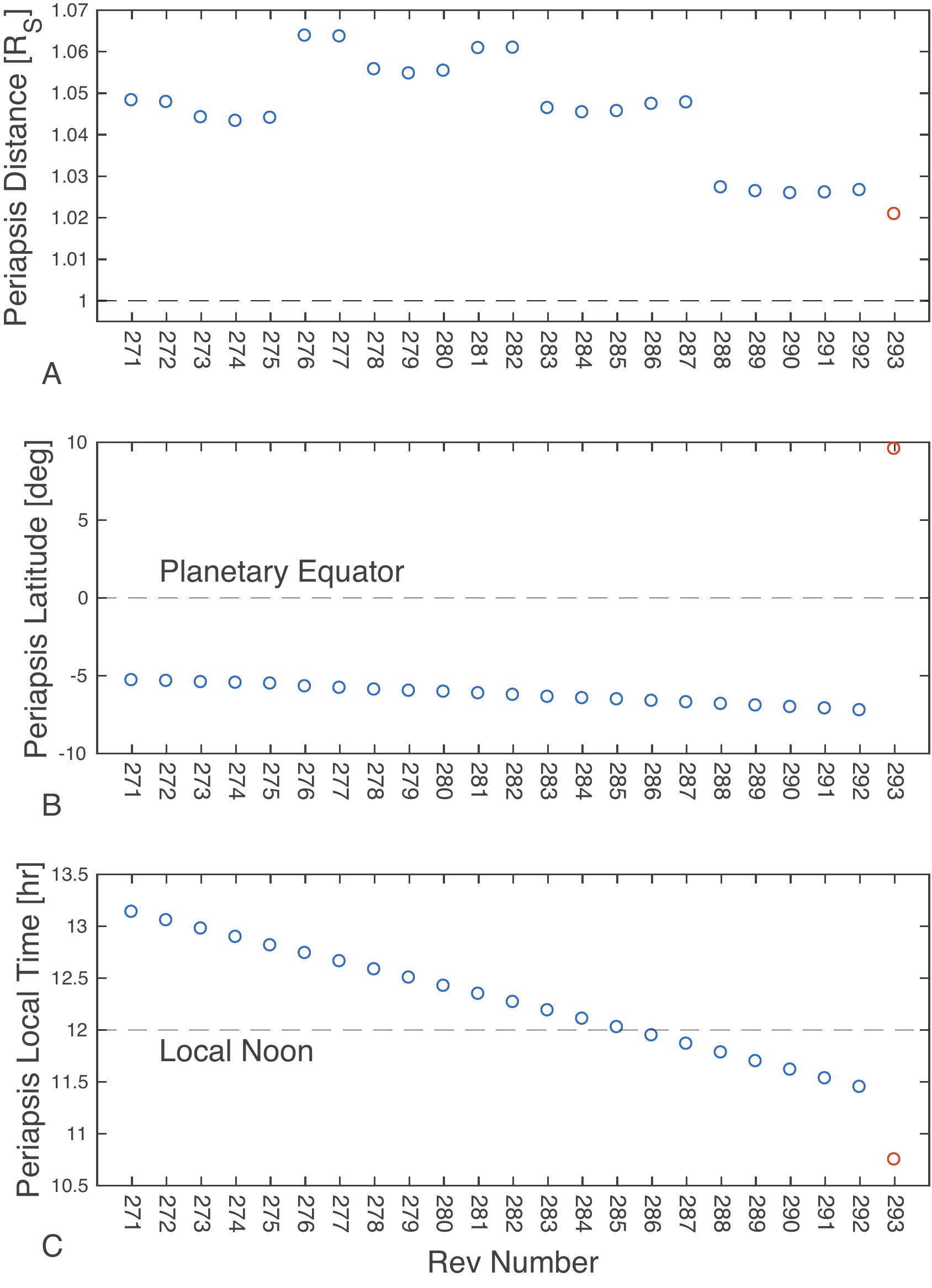}
\caption{Characteristics of the trajectory of Cassini Grand Finale orbits. Panel A shows the periapsis distance from the center of Saturn, panel B shows the periapsis latitude while panel C shows the periapsis local time as a function of the orbit (Rev) number.}
\label{fig2}
\end{figure}

The Grand Finale phase of the Cassini mission consists of 22.5 highly inclined, close-in orbits around Saturn between Apr 23rd 2017 (apoapsis time of first Grand Finale orbit Rev 271) and Sep 15th of 2017 (periapsis time of the last orbit Rev 293). Each Grand Finale orbit took $\sim$ 6.5 Earth days, with periapsis in the gap between Saturn and the inner edge of the D-ring and apoapsis near the orbit of Titan (Fig. \ref{fig1}). The trajectory and magnetic field measurements from selected Cassini Grand Finale orbits are shown in Figs. 1 - 4. Table 1 lists the periapsis information of all Cassini Grand Finale orbits including time, periapsis distance, altitude, latitude, and local time. Fig. \ref{fig1} shows the trajectory of a few typical Cassini Grand Finale orbits (the specific orbit shown in panel A is Rev 291, the ones shown in panels BC are Revs 271, 276, 292). The orbits featured inclination $\sim$ 62$^\circ$, the periapsis distance from the center of Saturn varied between 1.064 $R_S$ and 1.02 $R_S$ (1 $R_S$ = 60268 $km$), the periapsis latitudes were -6.2$^\circ$ $\pm$ 1$^\circ$ except that of the dive-in orbit which was $\sim$ 10$^\circ$, the periapsis local times were about $\pm$1 hour around local noon (Fig. 2).

\begin{table*}
\caption{Periapsis information of the Cassini Grand Finale orbits}
\centering
\begin{tabularx}{0.85\textwidth}{l c c c c c c}
\hline
 Rev  Num & Periapsis Date & UTC Time & Radial   & Altitude$^{a}$ & Latitude & Local Time  \\
                 &		           &		      &     Distance [$R_S$]	                 &  [$km$]             &  [deg]    &  [$hr$] \\
\hline
  271  & 26 Apr 2017		&	2017-116T09:03:34 & 1.048203 & 2963.16 & -5.296 & 13.135 \\ 
  272  & 02 May 2017  	&  	2017-122T19:42:15 & 1.047782 & 2939.30 & -5.364 & 13.054 \\
  273  & 09 May 2017   	&  	2017-129T06:16:39 & 1.044115 & 2719.79 & -5.429 & 12.974 \\
  274  & 15 May 2017  	&   	2017-135T16:45:20 & 1.043232 & 2667.90 & -5.486 & 12.894 \\
  275  & 22 May 2017  	&   	2017-142T03:14:28 & 1.043970 & 2713.47 & -5.535 & 12.812 \\
  276  & 28 May 2017   	&  	2017-148T14:26:22 & 1.063769 & 3910.65 & -5.717 & 12.738 \\
  277  & 04 Jun 2017   	&   	2017-155T01:42:28 & 1.063580 & 3901.04 & -5.793 & 12.659 \\
  278  & 10 Jun 2017  	&    	2017-161T12:53:15 & 1.055669 & 3427.14 & -5.907 & 12.581 \\
  279  & 17 Jun 2017   	&   	2017-167T23:55:43 & 1.054660 & 3367.97 & -5.974 & 12.501 \\
  280  & 23 Jun 2017   	&   	2017-174T10:57:42 & 1.055312 & 3409.05 & -6.047 & 12.422 \\
  281  & 29 Jun 2017   	&   	2017-180T22:14:15 & 1.060773 & 3740.92 & -6.160 & 12.345 \\
  282  & 06 Jul 2017  	&     	2017-187T09:35:23 & 1.060853 & 3747.78 & -6.239 & 12.266 \\
  283  & 12 Jul 2017  	&     	2017-193T20:48:00 & 1.046322 & 2875.56 & -6.366 & 12.185 \\
  284  & 19 Jul 2017  	&    	2017-200T07:54:43 & 1.045308 & 2816.82 & -6.456 & 12.104 \\
  285  & 25 Jul 2017		&      2017-206T18:59:19 & 1.045589 & 2835.97 & -6.539 & 12.024 \\
  286  & 01 Aug 2017 	&     	2017-213T06:09:10 & 1.047326 & 2943.12 & -6.632 & 11.945 \\
  287  & 07 Aug 2017 	&     	2017-219T17:23:16 & 1.047682 & 2967.09 & -6.725 & 11.864 \\
  288  & 14 Aug 2017  	&     	2017-226T04:23:03 & 1.027228 & 1737.60 & -6.826 & 11.779 \\
  289  & 20 Aug 2017  	&     	2017-232T15:23:00 & 1.026304 & 1684.73 & -6.924 & 11.696 \\
  290  & 27 Aug 2017  	&     	2017-239T02:18:10 & 1.025832 & 1659.24 & -7.026 & 11.613 \\
  291  & 02 Sep 2017 	&     	2017-245T13:13:00 & 1.026003 & 1672.41 & -7.126 & 11.531 \\
  292  & 09 Sep 2017  	&     	2017-252T00:09:45 & 1.026560 & 1709.06 & -7.229 & 11.448 \\
  293  & 15 Sep 2017 	&     	2017-258T10:31:41.755 & 1.020827 & 1443.63 & 9.559 & 10.749 \\
 \hline
 \multicolumn{7}{l}{$^{a}$Altitude here is defined as the minimum distance to the 1-bar spheroid with $a=60268 km$, $c=54364km$.}
\end{tabularx}
\end{table*}

\begin{figure}[h]
\centering
%
\includegraphics[width=0.475\textwidth]{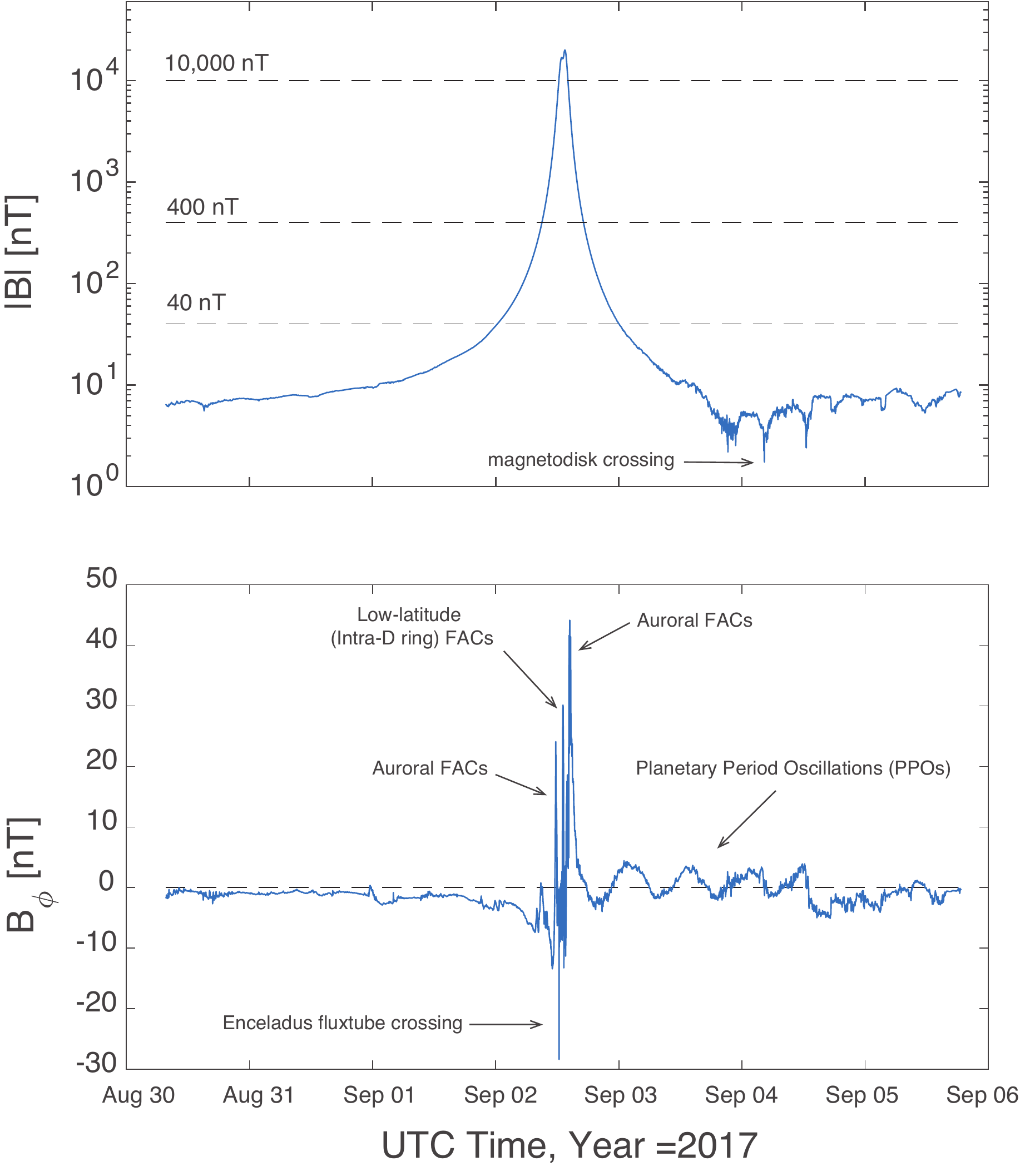}
\caption{Characteristics of the magnetic field measurements along a typical Cassini Grand Finale orbit from apoapsis to apoapsis (shown here is Rev 291). The top panel shows the total amplitude of the magnetic field, and the bottom panel shows the azimuthal component, which exhibits various magnetospheric features, including Auroral FACs, Intra-D ring FACs, Planetary Period Oscillations (PPOs), and Enceladus fluxtube crossing.}
\label{fig3}
\end{figure}

\begin{figure}[h]
\centering
%
\includegraphics[width=0.475\textwidth]{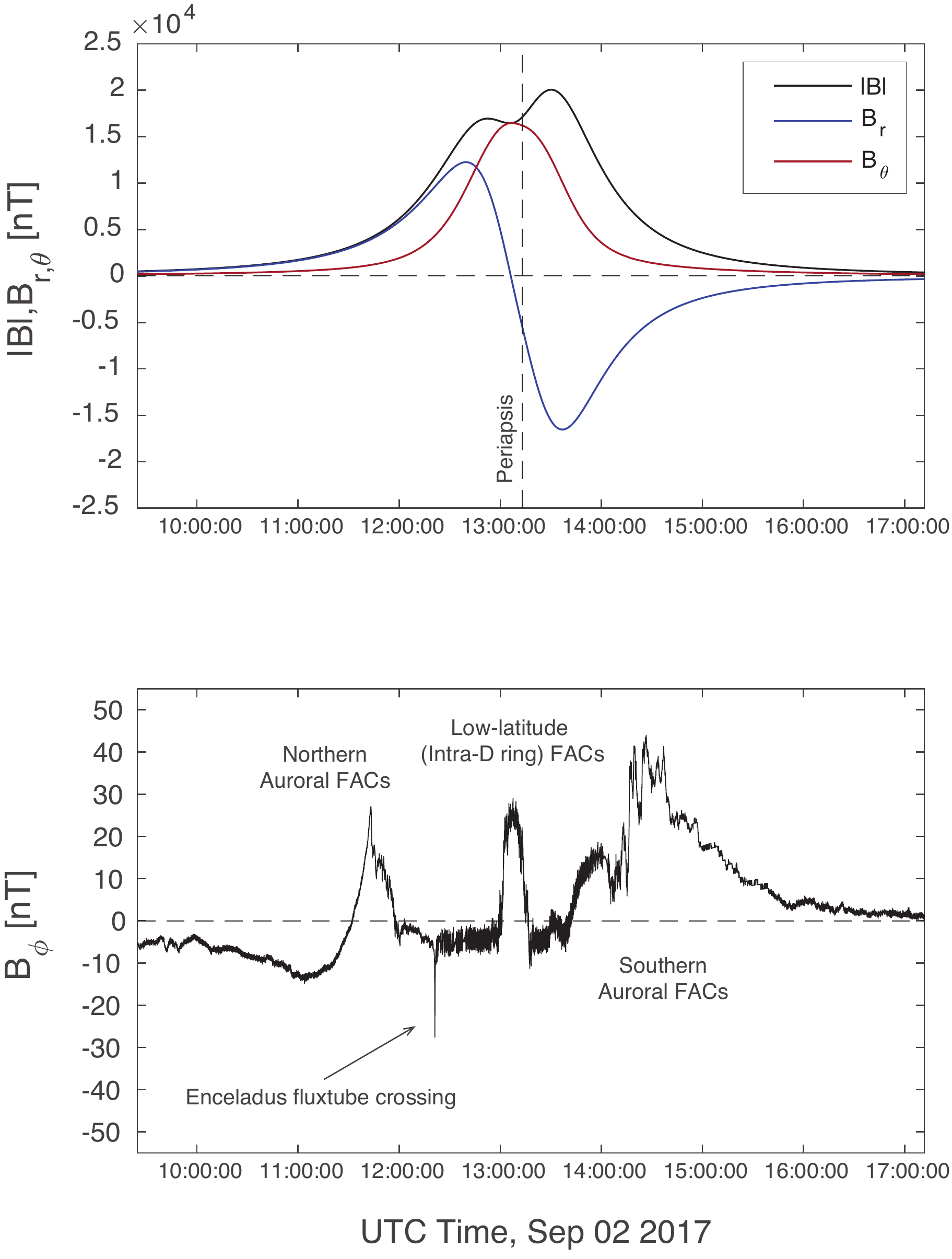}
\caption{Characteristics of the magnetic field measurements along a typical Cassini Grand Finale orbit within $\pm$ 4 hours around periapsis (shown here is Rev 291). The top panel shows the total amplitude of the magnetic field, the radial and meridional component, while the bottom panel shows the azimuthal component, which exhibits various magnetospheric features.}
\label{fig4}
\end{figure}
Fig. \ref{fig3} shows the measured magnetic field strength and the azimuthal component along one Cassini Grand Finale orbit, Rev 291, from apoapsis to apoapsis. It can be seen that the measured field strength ranges from $<$ 2 $nT$ to $>$ 20,000 $nT$. Thus, all four dynamical ranges of the fluxgate magnetometer \citep[]{Dougherty2004} were activated during a Cassini Grand Finale orbit. During the Grand Finale phase, the highest dynamical range of the fluxgate magnetometer, range 3, which can measure field above 10,000 $nT$ and up to 44,000 $nT$ with a digitization of 5.4 $nT$ were activated for the first time since the Cassini Earth Swing-by \citep[]{Southwood2001}. The minimum field strength along this orbit, 1.74 $nT$, was recorded during the crossing of the magnetodisk on the nightside (Fig. \ref{fig3}). 

To transform the vector magnetic field measurements from the spacecraft coordinate to an astronomical coordinate (e.g. the Saturn centered coordinate), the attitude of the spacecraft needs to be known to high precision. For example, the spacecraft attitude needs to be known to better than 0.25 milliradian (mrad) for the vector magnetic field to be known to within 5 $nT$ from the true values near the periapsis. The star tracker onboard Cassini was suspended intermittently during the Grand Finale orbits, which we refer to as Star ID suspensions. Table 2 lists the timing of the Star ID suspensions along each Grand Finale Orbit. The attitude of the spacecraft during the Star ID suspensions were reconstructed using information from the gyroscopes onboard \citep[see][for more information]{Burk2018}. Spacecraft rolls around two different spacecraft axes were designed and carried out along four Grand Finale orbits: Revs 272, 273, 284, 285. These spacecraft rolls enabled in-flight calibration of range 3 of the fluxgate magnetometer. The absolute scale of the fluxgate magnetometer was determined via comparing the simultaneous measurements carried out by the fluxgate magnetometer \citep[]{Southwood2001} and the helium magnetometer \citep[]{Smith2001} during the Earth Swing-by.

\begin{table*}
\caption{Star ID (SID) suspension time along the Cassini Grand Finale Orbits}
\centering
\begin{tabular}{l c c }
\hline
 Rev Num & First SID suspension & Second SID suspension \\
 \hline
  271  & 2017-116T08:35:19.000 to 09:54:57.854		&	None  \\ 
  272  & 2017-122T14:55:09 to 17:59:27 	&  	2017-122T18:19:40 to 20:53:22  \\
  273  & 2017-128T18:37:17 to 23:35:19   	&  	2017-129T03:53:09 to 08:51:11  \\
  274  & 2017-135T16:23:28 to 19:28:00  	&   	None  \\
  275  & 2017-142T02:52:31 to 05:57:03  	&   	None \\
  276  & 2017-148T13:54:12 to 16:37:24   	&  	None  \\
  277  & 2017-155T00:28:33 to 02:10:50  	&   	None \\
  278  & 2017-161T12:32:51 to 16:10:04  	&    	None  \\
  279  & 2017-167T23:47:12 to 168T01:09:29   	&   	None \\
  280  & 2017-174T10:37:32 to 14:14:45   	&   	None  \\
  281  & 2017-180T20:04:55 to 23:46:13   	&   	None  \\
  282  & 2017-187T09:06:11 to 10:03:09 	&     	None  \\
  283  & 2017-193T19:22:30 to 19:47:22  	&     	2017-193T20:13:51 to 22:21:18  \\
  284  & 2017-199T20:13:24 to 200T01:11:02  	&    	2017-200T05:30:20 to 10:27:58  \\
  285  & 2017-206T14:12:54 to 17:17:12		&      2017-206T17:38:06 to 20:18:02  \\
  286  & 2017-213T05:27:28 to 06:57:03 	&     	None  \\
  287  & 2017-219T15:51:03 to 16:20:43 	&     	2017-219T16:48:11 to 18:43:33  \\
  288  & 2017-226T02:51:54 to 03:21:09 	&     	2017-226T04:12:29 to 2017-226T06:12:09  \\
  289  & 2017-232T15:06:47 to 15:52:50  	&     	None  \\
  290  & 2017-239T00:44:31 to 04:07:28  	&     	None \\
  291  & 2017-245T12:44:47 to 14:18:00 	&     	None  \\
  292  & 2017-251T23:43:37 to 252T02:06:37 	&     	None  \\
  293  & 2017-258T10:11:19 to End of Mission 	&     	None  \\
 \hline
\end{tabular}
\end{table*}

Fig. \ref{fig3}B shows the measured azimuthal component, $B_\phi$, along Rev 291 which remains within $\pm$ 50 $nT$ and exhibits various magnetospheric features including the auroral FACs \citep[]{Hunt2014,Hunt2015,Hunt2018}, low-latitude (intra-D ring) FACs \citep[]{DC2018, Khurana2018, Provan2019b, Hunt2019}, crossing of the Enceladus fluxtube \citep[]{Sulaiman2018a}, and PPOs \citep[]{Provan2019}. Fig. 4 shows the total amplitude and all three components of the measured field in the Saturn centered KRTP coordinate within $\pm$ 4 hours of the periapsis along Rev 291. KRTP is a right-handed spherical polar coordinate, with its origin at the center of mass of Saturn, the polar axis (zenith reference) being the spin axis of Saturn, rotating at the IAU System III rotation rate of Saturn, while $r$, $\theta$, and $\phi$ {denote} radial, meridional, and azimuthal directions. The Enceladus fluxtube crossing, auroral FACs, and the intra-D ring FACs are better delineated in this zoomed-in version. The radial and meridional components exhibit a dipolar geometry, with $B_r$ being positive (negative) in the northern (southern) hemisphere while $B_\theta$ remains positive. The peak field strength is not encountered at the periapsis but at mid-latitude in the southern hemisphere. The overall features of the measured magnetic field are highly repeatable from orbit to orbit, although the magnetospheric features such as auroral FACs and intra-D ring FACs do exhibit orbit to orbit variations \citep[]{Provan2019b, Hunt2019}. 

\section{Saturn's magnetic equator position and its spatial variations}

The highly inclined nature of the Cassini Grand Finale orbits enabled direct determination of Saturn's magnetic equator positions, defined as where the cylindrical radial component of the magnetic field, $B_\rho$, vanishes. Fig. 5 displays the measured magnetic equator positions projected onto the $\rho-Z$ plane, where $\rho$ is distance from the spin-axis of Saturn and $Z$ is distance from the planetary equator of Saturn defined by the center of mass with northward being positive. Other than the Cassini Grand Finale measurements, the predictions from the Cassini 11 model \citep[]{DC2018} and the Cassini Saturn Orbital Insertion (SOI) measurements are shown in Fig. 5 as well. It can be seen that Saturn's magnetic equator is consistently displaced northward from the planetary equator. The measurements and the model predictions further demonstrate that the northward displacement of the magnetic equator, $Z_{MagEq}$, is not constant but varies as a function of $\rho$. Along the Grand Finale orbits where $\rho \sim 1.05 R_S$, the displacement is $\sim$ 2820 $km$ (0.0468 $R_S$). Along SOI, the spacecraft crossed the magnetic equator twice near $\rho \sim 2. 5 R_S$, where the displacement of the magnetic equator is $\sim$ 2300 $km$ (0.0382 $R_S$). The data-model comparison strongly suggests the axisymmetric part of the internal magnetic field is responsible for the majority of the observed spatial variations in $Z_{MagEq}$.

\begin{figure}[h]
\centering
%
\includegraphics[width=0.475\textwidth]{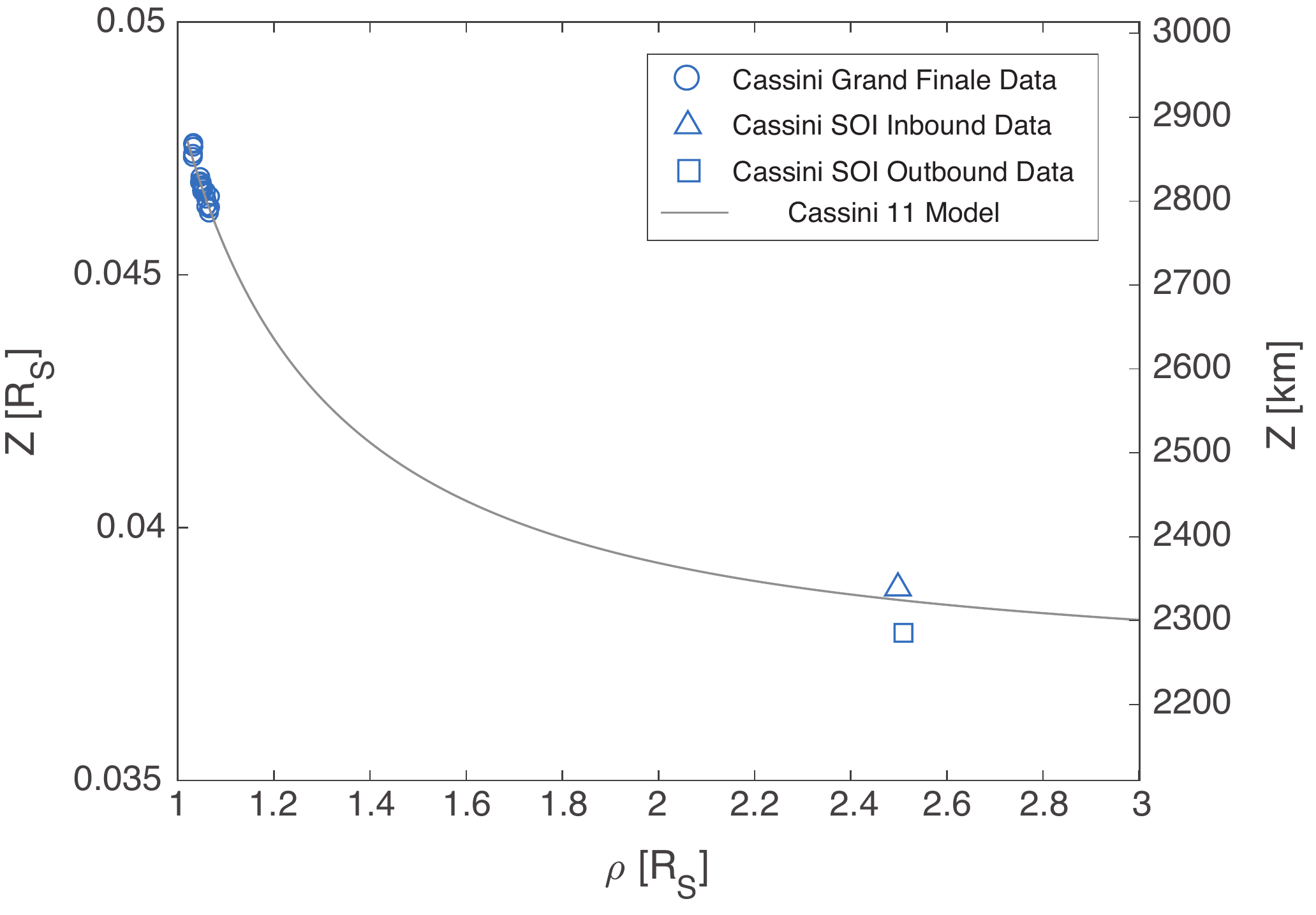}
\caption{Saturn's magnetic equator positions, defined as where the cylindrical radial component of the field vanishes ($B_\rho=0$), as measured along the Cassini Grand Finale orbits and the Cassini Saturn Orbital Insertion (SOI). The expected magnetic equator position based on the axisymmetric Cassini 11 model is over-plotted using the grey trace. It can be seen that Saturn's magnetic equator position varies as a function of distance from the spin-axis. The Cassini 11 model under predicts the measured magnetic equator positions by about 20 $km$ near $\rho=1.035$, the closest sets of measurements to the spin-axis. }
\label{fig5}
\end{figure}

\begin{figure}[h]
\centering
%
\includegraphics[width=0.475\textwidth]{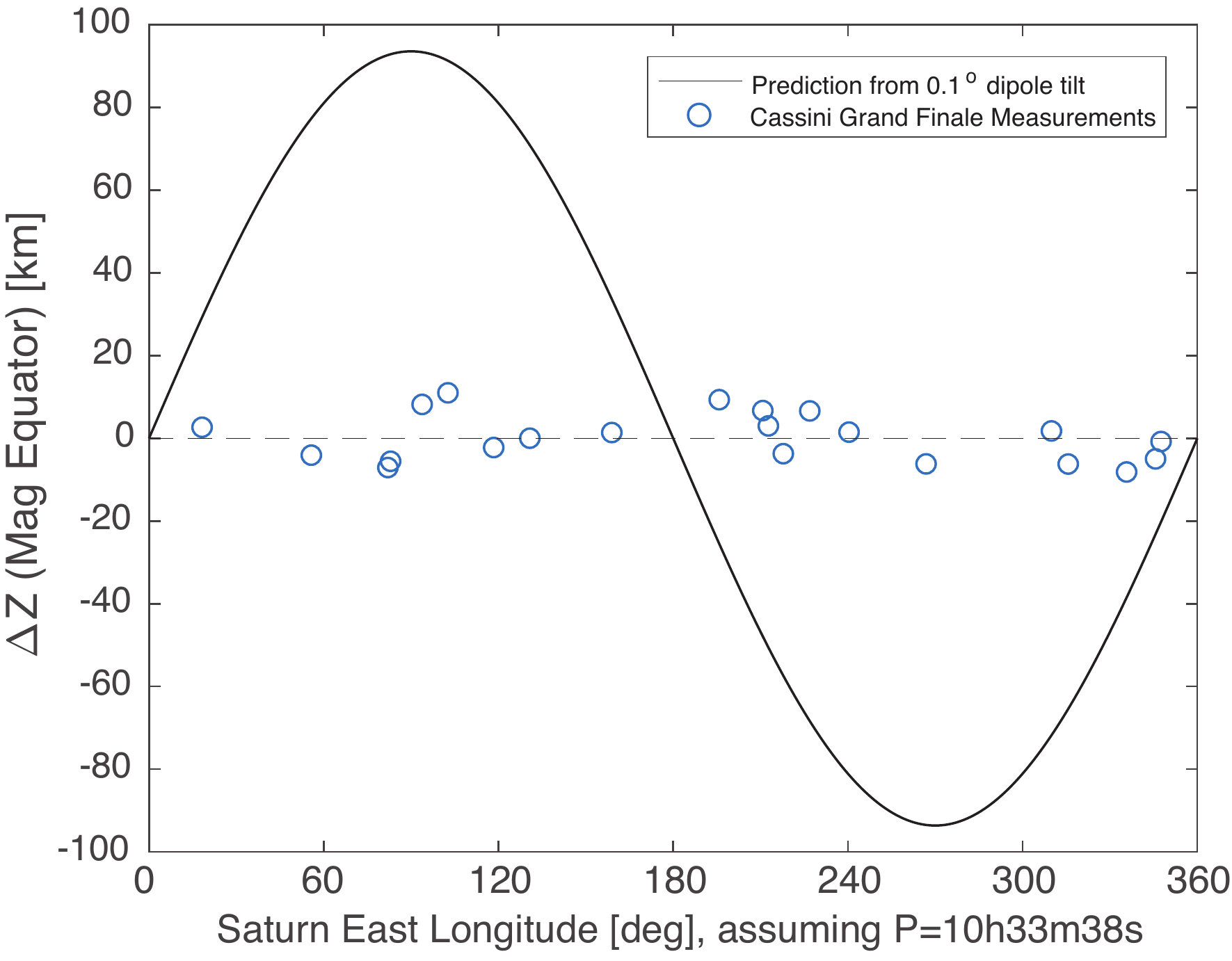}
\caption{Variations of Saturn's magnetic equatorial positions as a function of longitude compared. Prediction from a $0.1^\circ$ dipole tilt is over-plotted using the black trace. A degree-5 polynomial fitting, $Z_{MagEq}$ [$R_S$] =$0.215932/\rho^5-0.600580/\rho^4+0.651408/\rho^3-0.331803/\rho^2+0.084854/\rho+0.0291700$, in which $\rho$ is also in the unit of [$R_S$], has been removed from the measured magnetic equator positions. }
\label{fig6}
\end{figure}

\begin{figure}[h]
\centering
%
\includegraphics[width=0.475\textwidth]{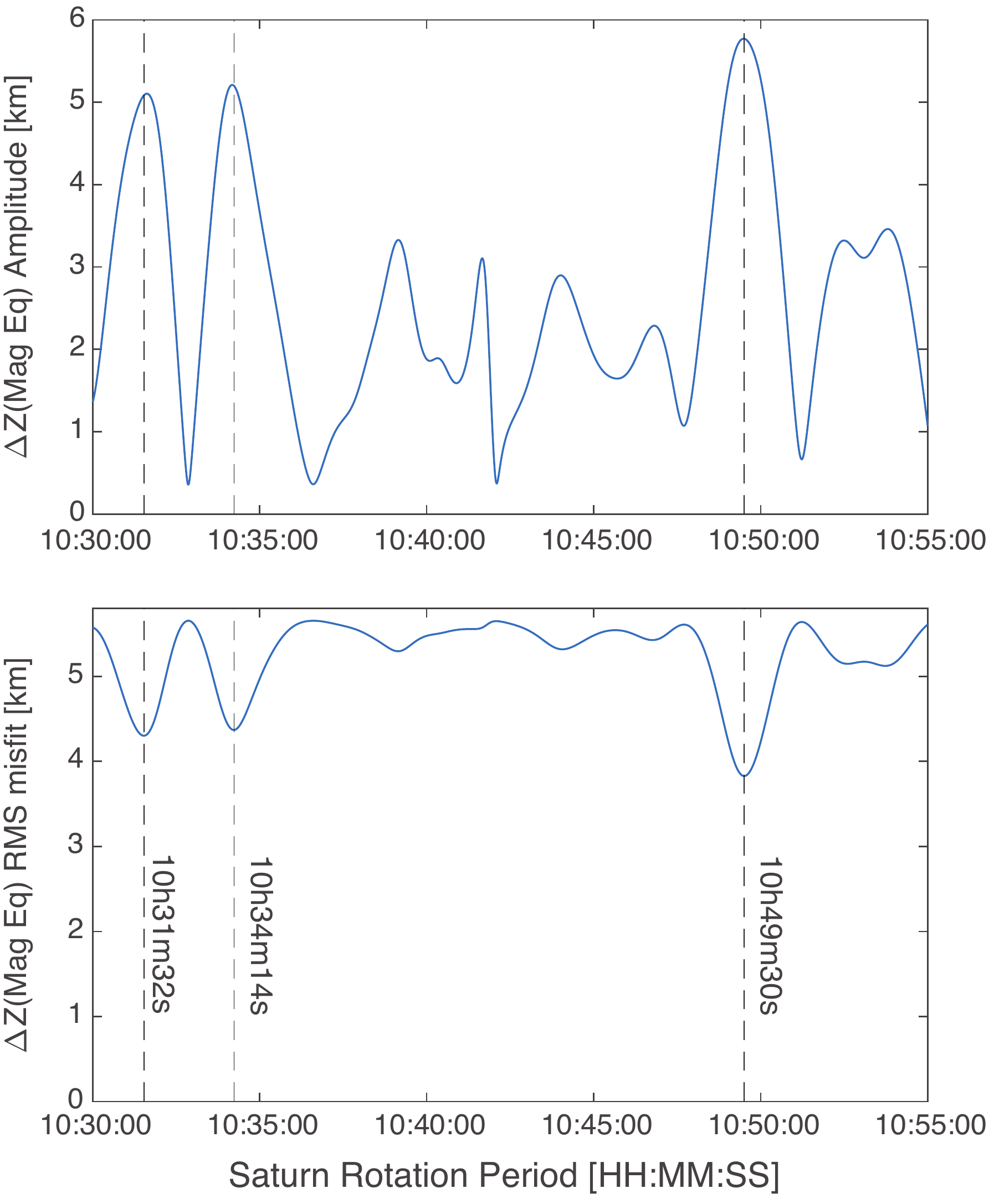}
\caption{Amplitude and root-mean-square (RMS) residual in searching for a $m=1$ pattern in Saturn's magnetic equator positions as a function of rotation rate of Saturn. Three dominant peaks are found at 10h49m30s, close to one of the planetary period oscillations period \citep[]{Provan2019}, 10h34m14s, close to the one of the ``internal" rotation rate derived from Saturn's 1-bar winds \citep[]{Read2009}, and 10h31m32s.}
\label{fig7}
\end{figure}

\begin{figure*}[t]
\centering
%
\includegraphics[width=0.95\textwidth]{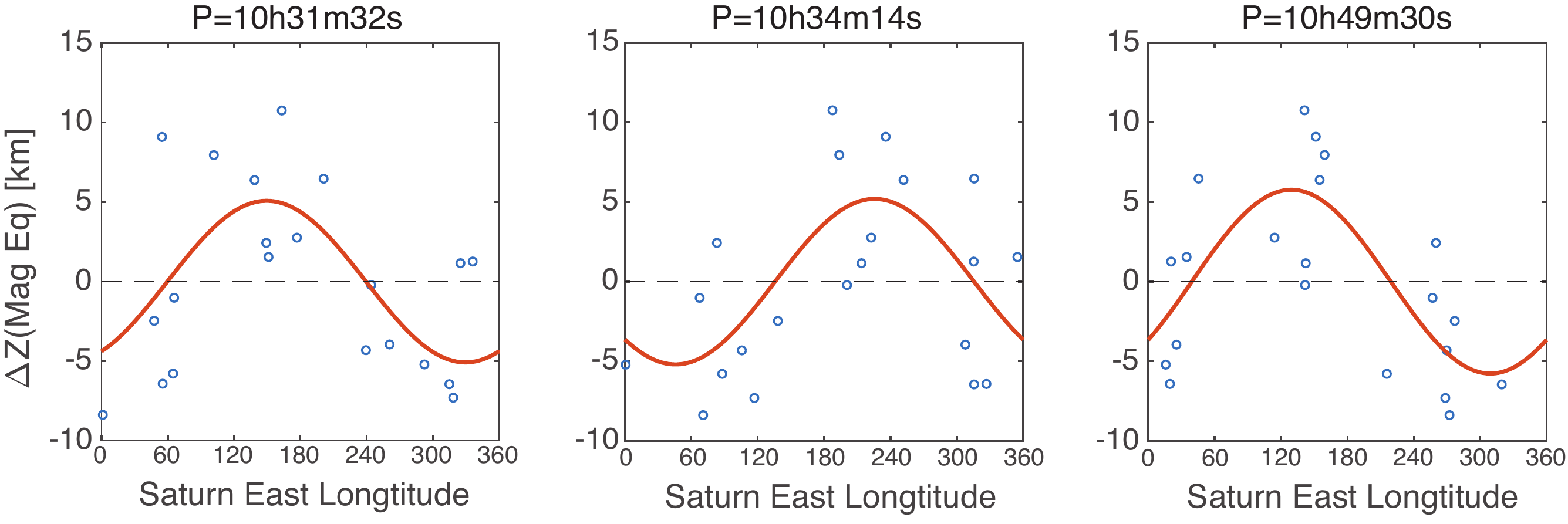}
\caption{Ordering of Saturn's magnetic equator positions as a $m=1$ pattern in longitude at three different rotation periods. }
\label{fig8}
\end{figure*}

In addition to the axisymmetric variations of $Z_{MagEq}$ with $\rho$,  multiple origins of perturbations in $B_\rho$ (e.g. the PPOs and non-axisymmetric internal magnetic moments such as $g_1^1$ and $h_1^1$) could cause additional $Z_{MagEq}$ variations. Near the magnetic equator crossing along the Grand Finale orbits, the relationship between the vertical displacement from the magnetic equator, $\Delta Z_{Mag Eq}= Z - Z_{Mag Eq}$, and $B_\rho$ can be approximated as 
\begin{equation}
\Delta Z_{Mag Eq} \quad [km] = 1.395 \quad [km/nT] \cdot B_\rho \quad [nT].
\end{equation}
Thus, a magnetic perturbation in $B_\rho$ {of} about 7.2 $nT$ would cause a displacement of the magnetic equator position by about 10 $km$. It should be noted that if such magnetic perturbations are of internal dipole origin (corresponding to $g_1^1$ and $h_1^1$), the corresponding $B_\phi$ would be about 3.6 $nT$. 

The measured peak-to-peak variations of $Z_{MagEq}$ at similar $\rho$ are less than 18 $km$ along the Grand Finale orbits. If the observed variations are caused by the internal non-axisymmetric dipole moments, the corresponding dipole tilt would be less than 0.01$^\circ$. {A dipole tilt much larger than 0.01 degrees can be safely ruled out by the data (Fig. \ref{fig6}).}

Here we carried out an explicit search for {m=1} non-axisymmetric patterns in the measured magnetic equator positions in addition to the variations with $\rho$. We first removed a degree-5 polynomial fit of the measured $Z_{MagEq}$ with {$1/\rho$}:

\begin{equation}
\begin{aligned}
Z_{MagEq} (\rho)= & 0.215932/\rho^5-0.600580/\rho^4 \\
			     & + 0.651408/\rho^3 -0.331803/\rho^2\\
			     & +0.084854/\rho+0.029170,\\
\end{aligned}
\end{equation}

in which both $Z_{MagEq}$ and $\rho$ are in the units of $R_S$. {A degree-5 polynomial fit yields an adequate description of the mean position of the magnetic equator without introducing additional spatial variations.} Then we search for a $sin(\phi+\phi_0)$ pattern in the residual magnetic equator positions $\Delta Z_{MagEq}$ {(Fig. \ref{fig6})}. Here $\phi$ is the east longitude in the spherical polar Saturn centered coordinate with a certain fixed rotation rate. We searched the possible range of rotation periods from 10h30m00s to 10h55m00s. The results are presented in Figs. \ref{fig7} \& \ref{fig8}. Interestingly, we find that the residual magnetic equator position can be ordered into a $sin(\phi+\phi_0)$ pattern at three different rotation periods, 10h31m32s, 10h34m14s, and 10h49m30s. The period 10h34m14s is almost identical to the internal rotation period of Saturn derived by \citet[]{Read2009} by considering the Arnol'd second stability criterion with the observed wind profile on Saturn. The ``best" ordering, judged by the amplitude of the pattern and the root-mean-square (RMS) residual, is at a period of 10h49m30s, close to the dominant northern PPO period - strangely no sign of southern PPO period \citep[]{Provan2019}. It should be noted that the peak amplitude of the $sin(\phi+\phi_0)$ pattern is less than 6 $km$ (thus the peak-to-peak variation is less than 12 $km$), translating into a dipole tilt of 0.0065$^\circ$ only. 

We will return to the search for internal non-axisymmetry with explicit modeling of the non-axisymmetric magnetic moments based on the vector magnetic field measurements in section {8}. The analysis so far has established that Saturn's internal magnetic field is exceptionally axisymmetric. 

\section{The sensitivity of Cassini Grand Finale MAG measurements to Saturn's internal magnetic field at depth}

Before proceeding to build models of Saturn's internal magnetic field from the Grand Finale MAG measurements, we first utilize the Green's function to forward calculate the sensitivity of the Grand Finale MAG measurements to Saturn's internal magnetic field at the ``dynamo surface", adopted as {the $a=0.75$ $R_S$, $c=0.6993$ $R_S$ isobaric ellipsoid} here. Estimation of the local magnetic Reynolds number $Rm$ guided the choice of dynamo surface for Saturn. Local $Rm$ is defined as $Rm=U_{conv}H_\sigma/\eta$, here $U_{conv}$ is the convective velocity, $H_\sigma=\left| \sigma/\frac{d\sigma}{dr}\right|$ is the conductivity scale-height, $\eta=1/\mu_0\sigma$ is the local magnetic diffusivity, where $\mu_0$ is the magnetic permeability and $\sigma$ is the local electrical conductivity. { According to the Saturn interior electrical conductivity model of \citet[]{Liu2008},} local $Rm$ reaches order 1 (10) at this depth if the convective velocity is on the order of 1 $mm/s$ ($cm/s$). Thus, downward continuation of the potential field to this depth seems appropriate. {Downward continuation of the potential field from the surface to certain depth inside the planet is only valid when there are no toroidal electrical currents in-between. Thus, downward continuation to depth much deeper than the $a=0.75$ $R_S$ isobaric surface cannot be guaranteed since local dynamo action is expected to become important around this depth.} Viewing the {downward continued} internal field around this depth would be most relevant for deciphering internal dynamics. 

Due to the highly axisymmetric nature of Saturn's internal magnetic field, the Green's function can be integrated in the azimuthal direction first and the mapping from the field at depth to the measurements along the spacecraft trajectory reduces to
\begin{equation}
B^{obs}_{r,\theta,\phi}(r,\theta)=\int_0^{\pi} B_r^{r_D}(\theta')\bar{G}_{r,\theta,\phi}(\mu)\sin\theta'd\theta'
\end{equation}
where the overbar denotes azimuthal integration. It can be easily shown that $\bar{G}_{\phi}=0$: axisymmetric current-free magnetic field has no azimuthal component. 

\begin{figure*}[htb]
\centering
%
\includegraphics[width=0.95\textwidth]{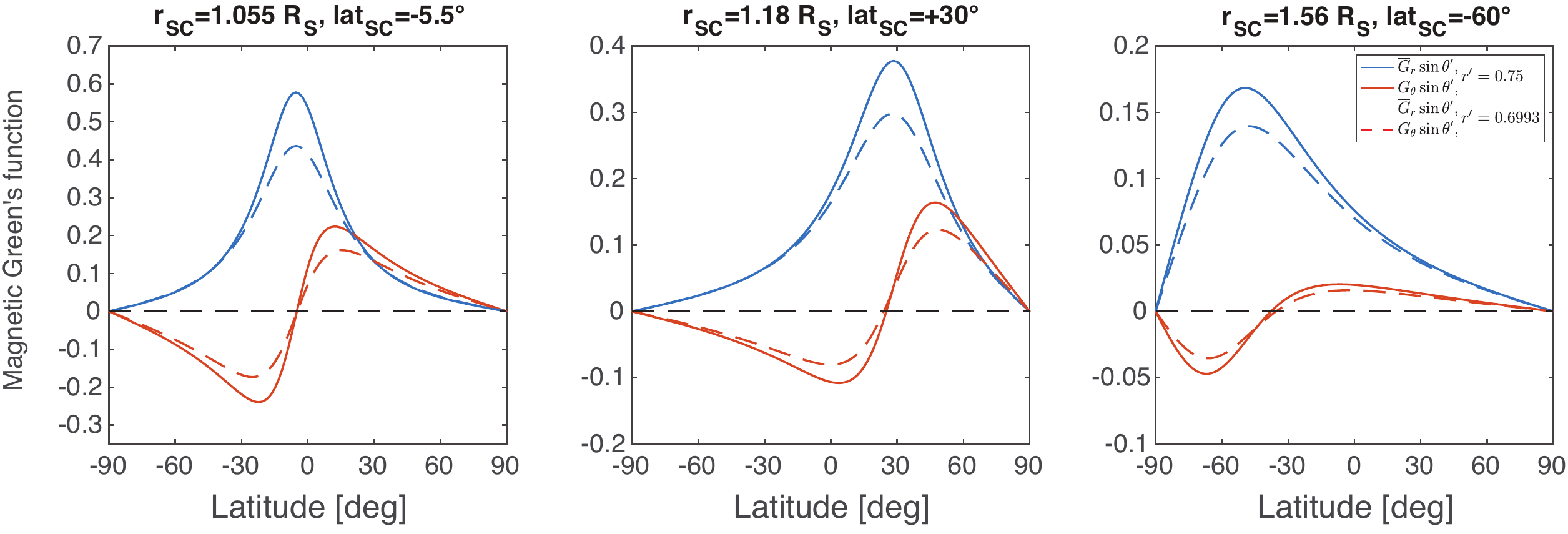}
\caption{Area-weighted, azimuthally integrated Green's function for Saturn's axisymmetric internal magnetic field evaluated at three different locations along a typical trajectory of Cassini Grand Finale orbits. {The solid traces show the Green's function with $r_D=0.75 R_S$, while the dashed traces show the Green's function with $r_D=0.6993 R_S$}.}
\label{fig9}
\end{figure*}

{Instead of switching to the confocal ellipsoidal coordinates to re-derive the Green's function, here we simply compute the Green's function for two different spherical surfaces, $r'=0.75$ $R_S$ and $r'=0.6993$ $R_S$, which bracket the $a=0.75$ $R_S$ isobaric surface. Qualitatively, the Green's function for the $a=0.75$ $R_S$ isobaric surface is expected to be close to $G_{r,\theta}^{0.75 R_S}$ near the equator and approach $G_{r,\theta}^{0.6993 R_S}$ towards the poles.} Fig. \ref{fig9} shows the azimuthally-integrated, area-weighted Green's function, $\bar{G}_{r,\theta}\sin\theta'$, for three locations along a typical Cassini Grand Finale trajectory (these locations are marked with blue crosses in Fig. \ref{fig1}B), which illustrates the sensitivity of the MAG measurements to Saturn's internal magnetic field at depth. 

{Taking the Green's function at the $r'=0.75$ $R_S$ surface for example,} at periapsis along the trajectory (Fig. \ref{fig9}A), $B_r^{obs}$ is mostly sensitive to $B_r^{0.75 R_S}$ around similar latitude ($-5^\circ$) with a half-amplitude-half-width (HAHW) of $\sim$ 20 degrees in latitude. On the other hand, $B_\theta^{obs}$ is mostly sensitive to $B_r^{0.75 R_S}$ at -22$^\circ$ and +12$^\circ$ latitude. At mid-latitude ($30^\circ$) along the trajectory, $B_r^{obs}$ is mostly sensitive to $B_r^{0.75 R_S}$ at similar latitude ($28.5^\circ$) with HAHW of 25 degrees, while $B_\theta^{obs}$ is mostly sensitive to $B_r^{0.75 R_S}$ at $4^\circ$ and $47^\circ$ latitude (Fig. \ref{fig9}B).  At high latitude ($-60^\circ$) along the trajectory,  $B_r^{obs}$ is mostly sensitive to $B_r^{0.75 R_S}$ at somewhat lower latitude ($-50^\circ$) with good sensitivity until $-80^\circ$ latitude, while $B_\theta^{obs}$ is most sensitive to $B_r^{0.75 R_S}$ around $-67^\circ$ with good sensitivity until $-80^\circ$ and even higher latitude (Fig. \ref{fig9}C). It should be noted that $\bar{G}_{r,\theta}\sin\theta'$ is always zero at the poles due to the area factor $\sin\theta'$. 

This forward calculation illustrates that MAG measurements along the Cassini Grand Finale trajectory are sensitive to Saturn's magnetic field at {depth} to very high latitudes ($\pm 80^\circ$). However, the Green's function is fairly wide in latitude near the polar region. This indicates that although the large-scale magnetic field at high-latitude should be well determined, the small-scale magnetic field beyond $60^\circ$ may not be uniquely determined. 

\section{Saturn's internal magnetic field from the Cassini Grand Finale MAG measurements}

Now we move on to retrieve Saturn's internal magnetic field from the Grand Finale MAG measurements. Although the Gauss coefficients are convenient mathematical tools to describe the magnetic field outside their source region, {the physical quantity} is the profile of Saturn's internal magnetic field at the dynamo surface and at the planetary surface. If there exist spatially localized features in the magnetic field near the spacecraft trajectory (e.g. a magnetic spot or a latitudinal flux band near the equator), the physical magnetic features could be well resolved by the MAG measurements yet the Gauss coefficients needed to represent the features might be uncertain and non-unique. This is because the Gauss coefficients are defined with respect to global functions which also depend on the field elsewhere on the globe. Thus, uncertainties and uniqueness of the solution should be evaluated in real space (e.g. evaluating the uncertainties and uniqueness in retrieved $B_r$ at the dynamo surface) rather than in the Gauss coefficients space, in particular when there is incomplete or uneven spatial coverage. 

In addition to the internal magnetic field generated by the MHD dynamo process in the deep interior, three categories of physical sources contribute to the MAG measurements along the spacecraft trajectory: magnetospheric currents (e.g. magnetodisk, magnetopause, magnetotail currents, and FACs), ionospheric currents, and electromagnetic induction response from the interior of Saturn. Along the close-in part of the trajectory (e.g. $r< 2.5 R_S$), magnetospheric contributions other than those from the adjacent FACs would appear as an external field and can easily be separated from the internal field given their different radial dependence. Moreover, existing analytical formulas for the magnetodisk field \citep[]{CAN1983, GD2004} allow a physics-based modeling. The magnetodisk field can be well approximated by a uniform $B_Z$ field around 12 $nT$ \citep[]{Bunce2007} along the closest part of the Grand Finale orbits. 

The ionospheric contributions, however, will appear as ``internal" field in the MAG measurements since the main conducting layer of the ionosphere, estimated to be $\sim$ 1100 $km$ above the 1-bar level \citep[]{MW2006}, lies below the trajectory of the Cassini Grand Finale orbits. Given the highly variable nature of Saturn's ionosphere from radio occultation and in-situ measurements \citep[]{Kliore2014,Wahlund2018,Persoon2019}, we do not expect the ionospheric contributed magnetic field to be stable with time, which provides one way of separating ionospheric contributions from deep dynamo contributions. In addition, we have made explicit estimations of the amplitude and profile of ionospheric contributed magnetic field at the top of the ionosphere and along the Cassini trajectory (see Appendix C). We found that their biggest contribution is to the axial dipole, which could amount to 6 $nT$. Their contributions to Gauss coefficients beyond degree-3 are expected to be less than 2.5 $nT$ (see Table \ref{tab:gn0_Hall} in Appendix C). Their impact on determining the deep dynamo magnetic field of Saturn can thus be explicitly assessed. The magnetospheric and ionospheric field, in particular their time variations, will induce additional internal magnetic field by setting up eddy currents in the conducting layer inside Saturn. For a time-varying signal with frequency close to the rotational frequency of Saturn or the orbital frequency of the Cassini Grand Finale orbits, the induction response will occur around 0.86 $R_S$ given our current understanding of Saturn's interior electrical conductivity profile \citep[]{Liu2008, CS2017Icarus, DC2018}. We will present our search for the induced internal field from the time-varying magnetodisk field in section 6. 

We first average the original 32 $Hz$ MAG measurements using a 10-sec window, keeping in mind that the raw attitude information from Star Trackers or gyroscopes were obtained once every 4 seconds. The contributions from the magnetodisk current are then determined orbit-by-orbit with the analytical formula given in \citet[]{GD2004} as the basis function. The determination of the magnetodisk field utilizes only MAG measurements with total field strength between 400 $nT$ and 10000 $nT$, corresponding approximately to radial distance between 1.5 $R_S$ and 3.8 $R_S$. These measurements are less affected by the determination of the small-scale internal magnetic field, thus offering better separation of internal and external magnetic field. {Furthermore, only field amplitude were employed to derive the magnetodisk field, reducing the effects of high-latitude field aligned currents.} Table 3 lists the parameters of the magnetodisk field for each Grand Finale orbit, from a non-linear {least-square} fitting procedure {based on the Levenberg-Marquardt method \citep[]{Levenberg1944, Marquardt1963}}. The value of magnetodisk field at the equator of Saturn, $B_Z$, along each orbit is listed in Table 3 as well. It can be seen that the magnetodisk $B_Z$ field varied between {12 $nT$ and 15.4 $nT$} along the Grand Finale orbits.

\begin{table}
\caption{Parameters of the magnetodisk field and the corresponding surface $B_Z$ along the Cassini Grand Finale orbits. {Here $a$ and $b$ are the radial distance of the inner and outer edge of the magnetodisk from the center of Saturn respectively, $D$ is the vertical half thickness of the magnetodisk, and $\mu_0I$ is the surface current amplitude, see \citet[]{CAN1983, GD2004, Bunce2007} for more details. In our analysis, only $\mu_0I$ were varied while $a$, $b$, and $D$ were fixed, due to the insensitivity of the MAG measurements inside 3 $R_S$ to the later three parameters.}}
\centering
\begin{tabular}{l c c c c c}
\hline
 Rev Num & a & b & $\mu_0I$  & D & Surface $B_Z$ \\
                 &	[$R_S$]	           &		[$R_S$]      &     [$nT$]	                 &  [$R_S$]             &  [$nT$]    \\
\hline
  271  & 6.5	&	20 & 48.1 & 2.5 & 12.2 \\ 
  272  & 6.5  	&  	20 & 47.8 & 2.5 & 12.1\\
  273  & 6.5   	&  	20 & 57.4 & 2.5 & 14.5  \\
  274  & 6.5  	&   	20 & 49.2 & 2.5 & 12.4  \\
  275  & 6.5  	&   	20 & 60.9 & 2.5 & 15.4  \\
  276  & 6.5   	&  	20 & 53.8 & 2.5 & 13.6 \\
  277  & 6.5   	&   	20 & 48.2 & 2.5 & 12.2  \\
  278  & 6.5 	&    	20 & 54.8 & 2.5 & 13.9  \\
  279  & 6.5   	&   	20 & 51.2 & 2.5 & 12.9  \\
  280  & 6.5  	&   	20 & 47.7 & 2.5 & 12.1  \\
  281  & 6.5  	&   	20 & 57.0 & 2.5 & 14.4  \\
  282  & 6.5	&     	20 & 51.3 & 2.5 & 13.0  \\
  283  & 6.5  	&     	20 & 52.7 & 2.5 & 13.3  \\
  284  & 6.5  	&    	20 & 51.0 & 2.5 & 12.9  \\
  285  & 6.5	&      20 & 56.5 & 2.5 & 14.3 \\
  286  & 6.5 	&     	20 & 56.9 & 2.5 & 14.4  \\
  287  & 6.5 	&     	20 & 55.3 & 2.5 & 14.0  \\
  288  & 6.5  	&     	20 & 57.5 & 2.5 & 14.5 \\
  289  & 6.5  	&     	20 & 60.5 & 2.5 & 15.3  \\
  290  & 6.5  	&     	20 & 59.3 & 2.5 & 15.0  \\
  291  & 6.5 	&     	20 & 56.4 & 2.5 & 14.2  \\
  292  & 6.5  	&     	20 & 57.2 & 2.5 & 14.5  \\
  293  & 6.5 	&     	20 & 47.6 & 2.5 & 12.0  \\
 \hline
\end{tabular}
\end{table}

\subsection{Inversion of Saturn's axisymmetric internal magnetic field with Gauss coefficients representation}

After removal of the magnetodisk field, we solve for Saturn's axisymmetric internal magnetic field with the traditional Gauss coefficients representation first. Since we are only seeking an axisymmetric internal field solution at this step, which has zero contribution to the azimuthal field $B_\phi$, only $(B_r,B_\theta)$ from the measurements were adopted. Excluding $B_\phi$ has no {effect} on the model solutions but does affect the values of the reported RMS residual.  

We tested two different data selection (DS) criteria: 1) only selecting measurements with $|B| > 10000 nT$, which approximately corresponds to $r < 1.5 R_S$ along the Grand Finale orbits; 2) selecting all measurements with $r < 3 R_S$, which approximately corresponds to $|B| > 1274 nT$. Criterion 1 avoids measurements during the crossing of the high latitude FACs \citep[]{DC2018} whilst criterion 2 extends the data to the maximum latitude coverage. 

\subsubsection{Un-regularized inversion}

The forward linear problem can be formulated as 
\begin{equation}
\mathbf{data} = G \: \mathbf{model},
\end{equation}
in which $\mathbf{data}$ represents MAG measurements with the magnetodisk field removed, $\mathbf{model}$ represents the Gauss coefficients, and $G$ represents the matrix expression of equation (\ref{eqn:B_gh}). In un-regularized inversion, we seek to minimize the data-model difference 
\begin{equation}
\left| \mathbf{data} - G \: \mathbf{model} \right|^2,
\end{equation}
without placing explicit constraints on the behavior of the model. 

We monotonically increase the maximum spherical harmonic (SH) degree, $n_{max}$, of the axisymmetric internal field model and examine the behavior of the data-model fit. Both the RMS residual and the vector residual at each data point are evaluated. This exercise aims at revealing the minimum spectral content required by the measurements. 

\begin{figure}[h]
\centering
%
\includegraphics[width=0.475\textwidth]{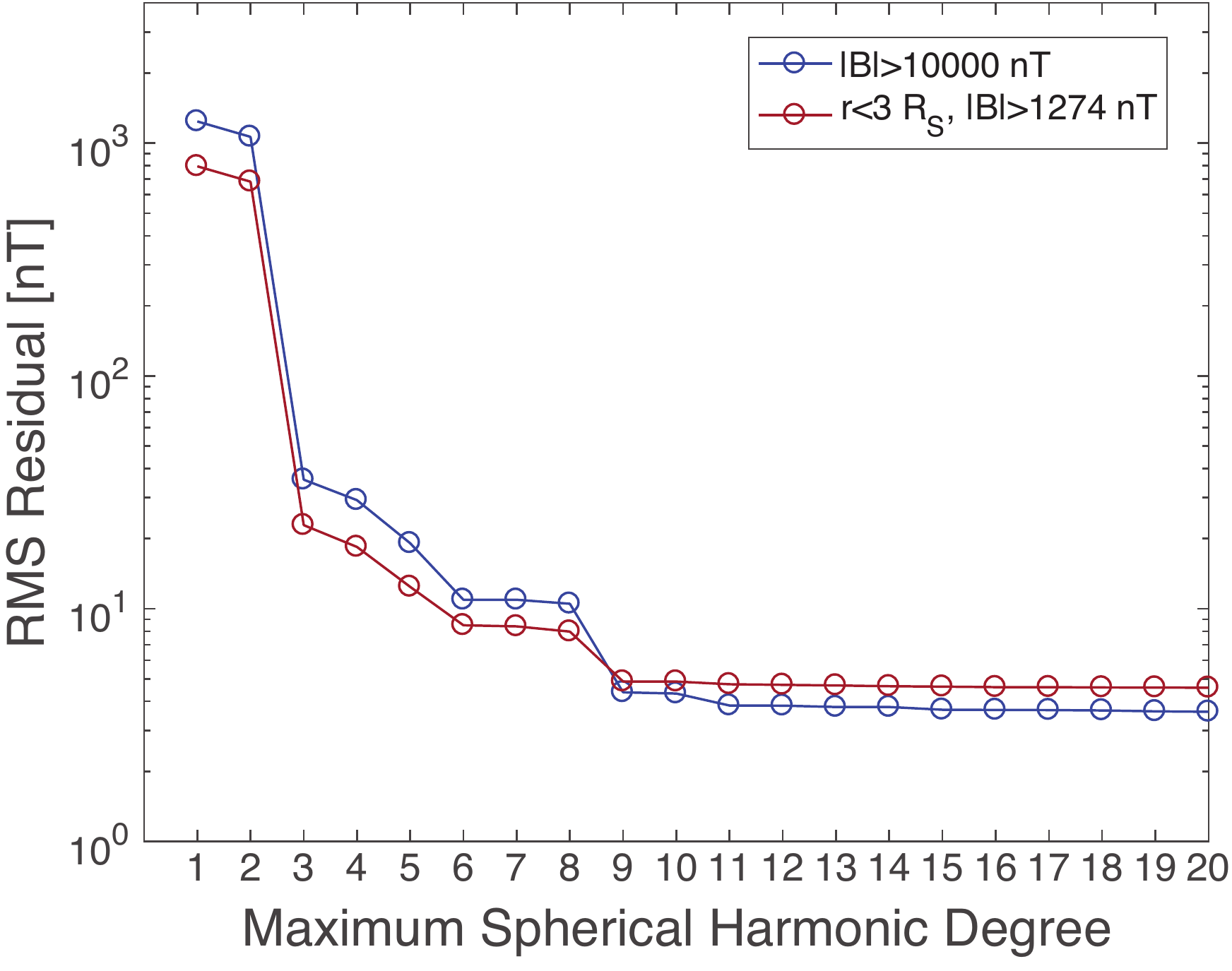}
\caption{Root-mean-square (RMS) residual from the un-regularized axisymmetric inversion. Only $(B_r, B_\theta)$ were adopted in this analysis. The two different traces represent two different data selection criteria.}
\label{fig10}
\end{figure}

Table 4 lists the Gauss coefficients from the un-regularized inversion with the two different data selection criteria, while Fig. \ref{fig10} shows the RMS residual. It can be seen that although the RMS residuals corresponding to the two different data selection criteria behave slightly differently, the resulted model solutions from the two data selection criteria are almost identical. This indicates the FACs do not have a significant impact on the internal field modeling given the Grand Finale trajectory. Table 4 also shows that the Gauss coefficients beyond degree 3 are on the order of 100 $nT$ or less, significantly smaller than those of degrees 1 - 3.

\begin{table}[h]

\caption{Gauss coefficients of the un-regularized inversion of Saturn's axisymmetric internal magnetic field with two different data selection (DS) criteria.}
\centering
\resizebox{\columnwidth}{!}{%
\begin{tabular}{r  r r r r r r }
\hline
   & $n_{max}=3$  & $n_{max}=3$ & $n_{max}=6$ & $n_{max}=6$ & $n_{max}=9$ & $n_{max}=9$ \\

  [$nT$] & DS 1  & DS 2 & DS 1 & DS 2 & DS 1 & DS 2 \\
\hline
  $g_1^0$		& 21120	&  21127	&  21156	& 21150	& 21139   	& 21139	\\
   $g_2^0$  	& 1522  	&  1527 	& 1591	&  1586	& 1578  	&  1576	 \\
   $g_3^0$  	& 2218   	&  2223 	& 2300   	&  2291	& 2255   	&  2255	 \\
   $g_4^0$  	&    		&  		& 116   	&  108 	& 82  	&  77		\\
   $g_5^0$  	&    		&  		& 77   	&   71	& $-9$  	&  $-9$	\\
   $g_6^0$  	&    		& 		& 49   	&   45	& $-3$   	& $-8$	\\
   $g_7^0$  	&    		&  		&   		&  		& $-100$  & $-100$	\\
   $g_8^0$  	&    		&  		&    		& 	 	& $-36$   	& $-39$ 	\\
   $g_9^0$  	&    		& 		&    		& 	 	& $-55$   	& $-54$	\\
 \hline
\end{tabular}
}
\end{table}

The RMS residual in the un-regularized inversion decreases monotonically with the maximum SH degree, with a few distinct features: 1) the RMS residual drops by more than an order of magnitude from $n_{max}=2$ to $n_{max}=3$, 2) the RMS residual remains roughly constant ($\sim$ 10 $nT$) between $n_{max}=6$ and $n_{max}=8$, 3) the RMS residual decreases by more than a factor of two from $n_{max}=8$ to $n_{max}=9$.

\begin{figure*}[htbp]
\centering
%
\includegraphics[width=0.90\textwidth]{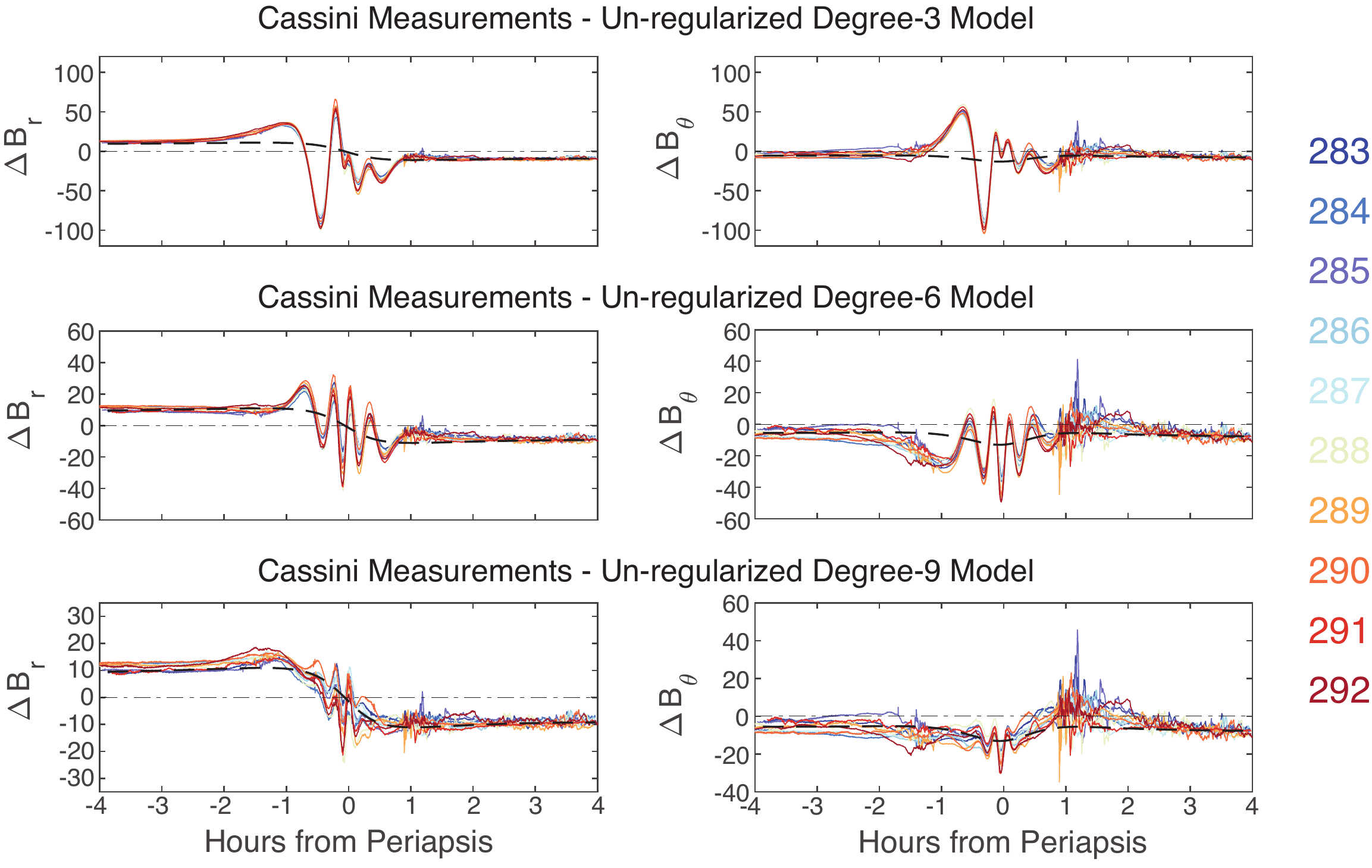}
\caption{Component residuals, $(\Delta B_r, \Delta B_\theta)$, from the un-regularized degree 3, degree 6, and degree 9 models along Rev 283 to Rev 292 within $\pm$ 4 hours of the periapsis. In each panel, thick black dashed line represents contribution from the mean magnetodisk field.}
\label{fig:dBrt_C369}
\end{figure*}

Fig. \ref{fig:dBrt_C369} shows the vector residuals as a function of time from periapsis along the S/C trajectory for Rev 283 to Rev 292, with the contribution from the mean magnetodisk field being over-plotted (thick black dashed lines). The behavior along all other orbits are quite similar. It can be seen that the vector residuals from the un-regularized degree-3 model feature larger amplitude and larger spatial-scale in the northern hemisphere while the vector residuals from the un-regularized degree-6 model features mostly north-south symmetric oscillations.  The residuals from the un-regularized degree-9 model are broadly consistent with the average magnetodisk field, except within [-20, +10] minutes around the periapsis. 

\begin{figure}[h]
\centering
%
\includegraphics[width=0.45\textwidth]{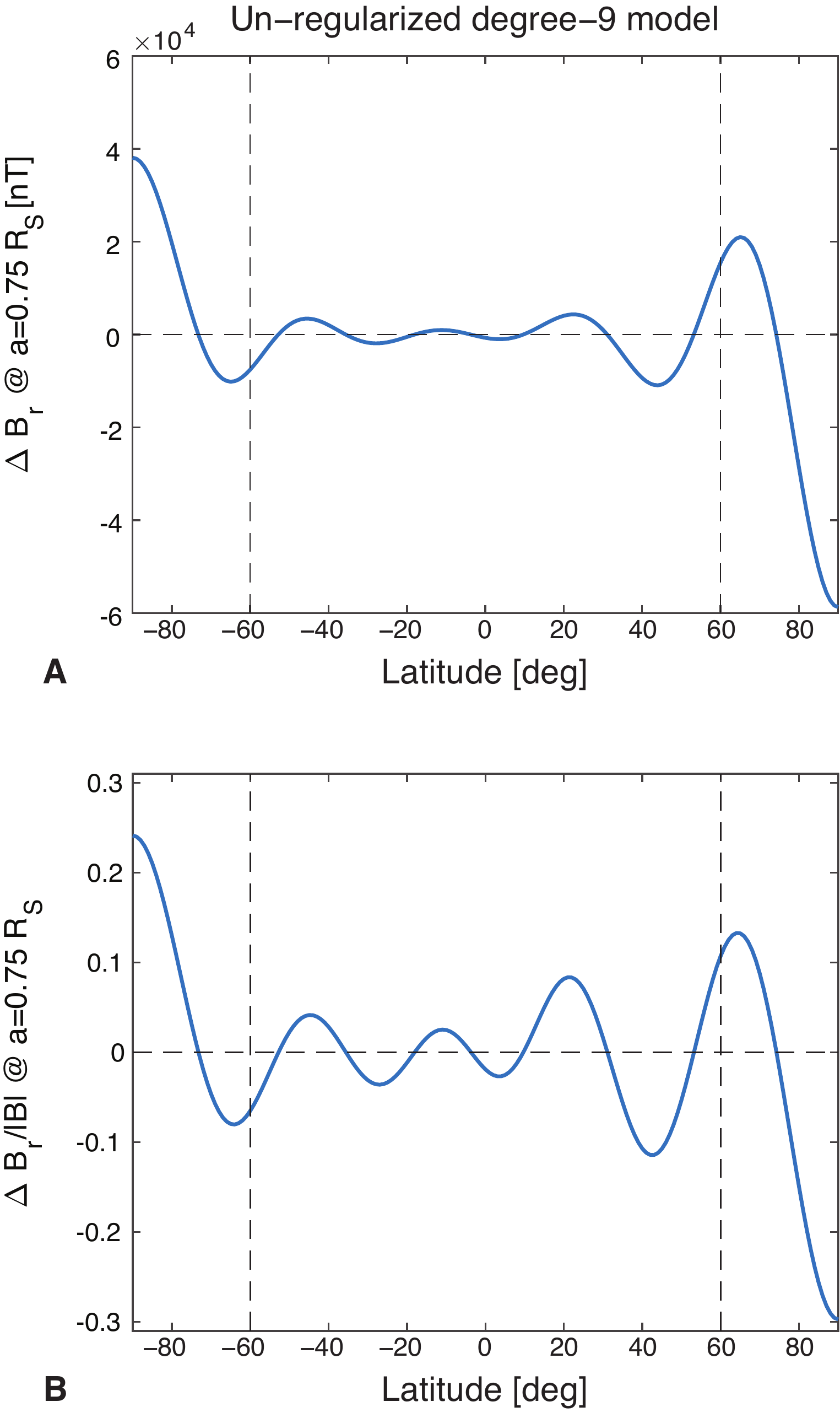}
\caption{Profile of the small-scale $(n>3)$ axisymmetric magnetic field $\Delta B_r$ and {$\Delta B_r (n>3)/|B(n\le3)|$ at the $a=0.75$ $R_S$, $c=0.6993$ $R_S$ isobaric surface} according to the un-regularized degree-9 model. It can be seen that in this un-regularized model, {$\Delta B_r$ above $\pm$60$^\circ$ latitude are about 3.75 times larger than that within $\pm$60$^\circ$, and $\Delta B_r/|B|$ above $\pm$60$^\circ$ are about 2.5 times larger than that within $\pm$60$^\circ$.}}
\label{fig:dBr_075_SVD9}
\end{figure}

Given that the un-regularized degree-9 model fits the measurements reasonably well except very close to the periapsis, why not take it as a new basis solution of Saturn's internal magnetic field? To answer this question, we examine the magnetic perturbations associated with Gauss coefficients above degree 3 at {the $a=0.75$ $R_S$, $c=0.6993$ $R_S$ isobaric ellipsoidal surface}. As shown in Fig. \ref{fig:dBr_075_SVD9}, when evaluated at {the $a=0.75$ $R_S$ isobaric surface}, {$\Delta B_r$ associated with the degree 4 - 9 coefficients of the un-regularized degree-9 model features 3.75 times higher values above 60$^\circ$ latitude compared to those within $\pm$ 60$^\circ$ latitude. Moreover, the fractional amplitude of the small-scale field perturbations $\Delta B_r (n>3)/|B(n\le3)|$ above 60$^\circ$ are about 2.5 times larger than that within $\pm$60$^\circ$.} Given that the Cassini spacecraft did not go much beyond $\pm$60$^\circ$ latitude during the Grand Finale phase, the model field behavior beyond $\pm$60$^\circ$ latitude is likely to be neither justified nor uniquely determined by the measurements. Thus, we turn to the regularized inversion technique \citep[]{HB1996, Gubbins2004} to construct internal field models for Saturn that not only fit the Cassini measurements but are also well-behaved. Here, we define ``well-behaved" in the sense that the {fractional} amplitude of the small-scale field perturbations beyond 60$^\circ$ are similar to that within $\pm$ 60$^\circ$. This definition of ``well-behaved" is a subjective choice, but it is a reasonable one given the available measurements.

\subsubsection{Regularized inversion}

In regularized inversion, in addition to seeking models that fit the data, constraints are placed on the behavior and properties of the model. This can be formulated as minimizing
\begin{equation}
\left| \mathbf{data} - G \: \mathbf{model} \right|^2 + \gamma^2 \left|L \: \mathbf{model} \right|^2,
\end{equation}
here $\gamma$ is a tunable damping parameter controlling the relative importance of model constraints and data-model misfit, while $L$ represents the particular form of constraint on the model. Here we seek to minimize the surface integrated power in the radial flux, $ \int B_r^2(n > 3) d\Omega$, at {$r=0.6993 R_S$. Since we expect the regularization to mainly constrain the behavior of the magnetic field above $\pm$60$^\circ$ latitude, we set the regularization radius to 0.6993 $R_S$, the polar radius of the $a=0.75$ $R_S$ isobaric surface}. Thus, the model constraint is
\begin{equation}
L=\frac{n+1}{\sqrt{2n+1}}\left( \frac{R_p}{r_{damp}} \right)^{n+2}
\end{equation}
for $n>3$ and $L=0$ for $n \le 3$, in which $R_p$ is the radius of the planet and {$r_{damp}$ is the damping} radius at which the constraints are placed. Here, $R_p = R_S$, and {$r_{damp}=0.6993$ $R_S$}.

\begin{figure}[htbp]
\centering
%
\includegraphics[width=0.475\textwidth]{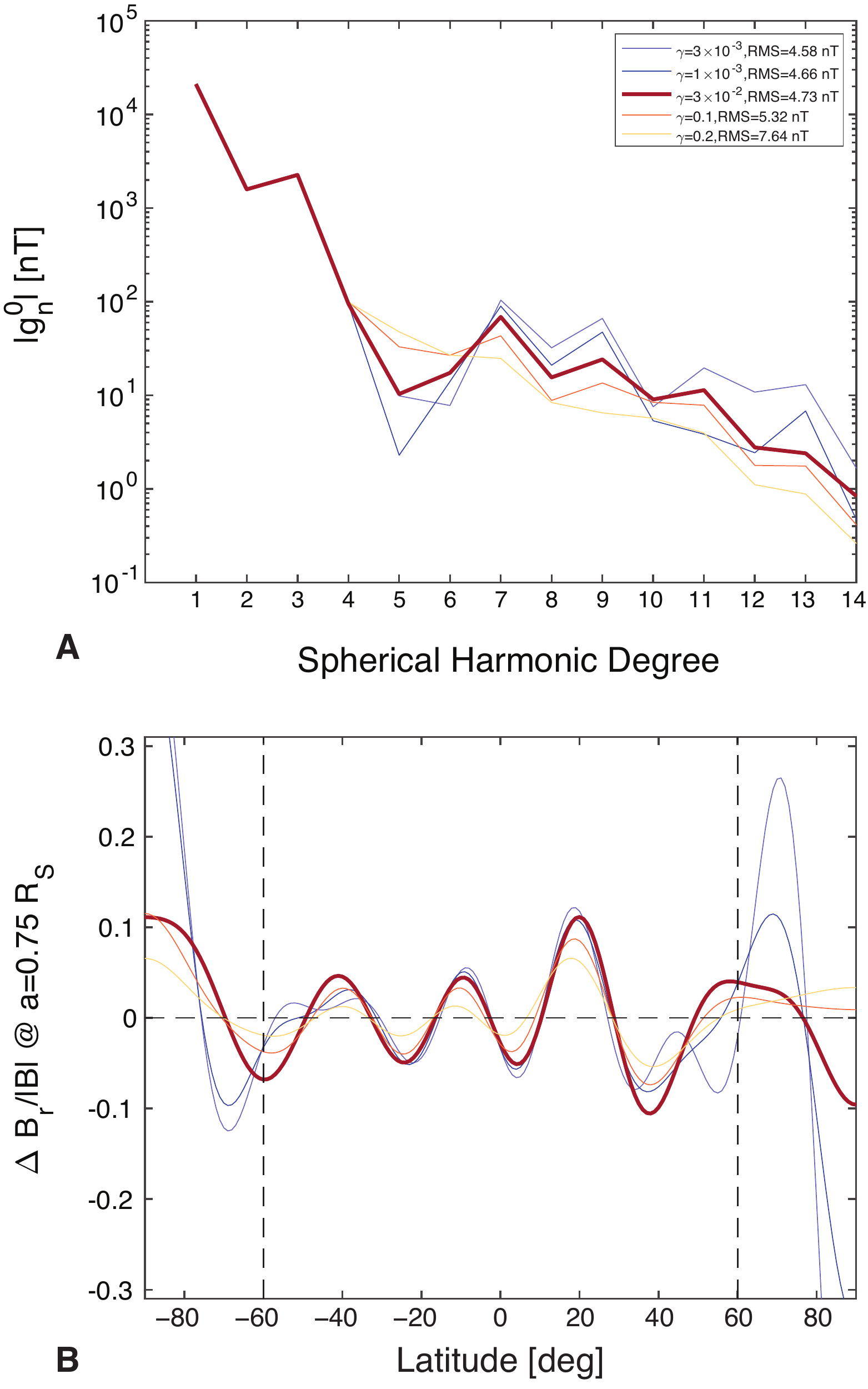}
\caption{Gauss coefficients and $\Delta B_r (n>3)/|B(n\le3)|$ { at the $a=0.75$ $R_S$, $c=0.6993$ $R_S$ isobaric surface} from a survey of regularized inversion based on Cassini Grand Finale MAG measurements. The thick red traces represent our preferred solution, {the Cassini 11+ model}. }
\label{fig:gn0_dBr_Reg}
\end{figure}

Fig. \ref{fig:gn0_dBr_Reg} displays the Gauss coefficients and {$\Delta B_r(n>3)/|B(n\le3)|$ at the $a=0.75$ $R_S$, $c=0.6993$ $R_S$ ellipsoidal surface} from a survey of regularized inversion with different damping parameters. The preferred solution is highlighted using thick red traces in both panels. Compared to the un-regularized degree-9 model, our preferred solution features {$\Delta B_r/|B|$} with similar amplitude beyond $\pm$60$^\circ$ and within $\pm$ 60$^\circ$. Moreover, Fig. \ref{fig:gn0_dBr_Reg} shows that the model $B_r$ are broadly similar within $\pm$ 60$^\circ$. 

This preferred solution constructed from the entire Grand Finale dataset is very similar to the Cassini 11 model \citep[]{DC2018} derived from 9 of the first 10 Grand Finale orbits in the profile of $B_r$ and in the Gauss coefficients (see Table 5 for the Gauss coefficients). We refer to this newly constructed model as the Cassini 11$+$ model.

\begin{table}[htb]
\caption{{Gauss Coefficients of newly derived Cassini 11+ model compared to that of the Cassini 11 model \citep[]{DC2018}}}
\centering
\begin{tabular}{r r r}
\hline
  [$nT$] & Cassini 11 & Cassini 11+  \\
\hline
  $g_1^0$  & 21140   & 21141 \\
   $g_2^0$  & 1581  & 1583 \\
   $g_3^0$  & 2260   & 2262 \\
   $g_4^0$  & 91   & 95 \\
   $g_5^0$  & 12.6   & 10.3  \\
   $g_6^0$  & 17.2   & 17.4 \\
   $g_7^0$  & $-59.6$ &  $-68.8$ \\
   $g_8^0$  & $-10.5$ &  $-15.5$ \\
   $g_9^0$  & $-12.9$ &  $-24.2$ \\
   $g_{10}^0$  & 15   & 9.0 \\
   $g_{11}^0$  & 18   & 11.3 \\
   $g_{12}^0$  &    & $-2.8$ \\
   $g_{13}^0$  &    & $-2.4$ \\
   $g_{14}^0$  &    & $-0.8$ \\
 \hline
\end{tabular}
\end{table}

\subsection{Inversion of Saturn's internal magnetic field with Green's function}

\subsubsection{The eigenvectors of the inverse problem formulated with Green's function}

In addition to the traditional Gauss coefficients representation, the inverse problem for the internal magnetic field can be formulated with Green's function representation. In this formulation, the $\mathbf{model}$ in 
\begin{equation}
\mathbf{data} = G \: \mathbf{model}
\end{equation}
is the profile of $B_r$ at the dynamo surface, and $G$ is the matrix expression of equation (\ref{eqn:B_Green}). {For simplicity, we choose $B_r$ at $r_d=0.6993$ $R_S$, same as the damping radius in our regularized inversion, as the $\mathbf{model}$ here.}  

Each {eigenvector} of the inverse problem is a profile of axisymmetric $B_r^{r_d}$ {as a function of latitude}, which we denote as $B_i^{r_d}$, here $i$ is the order of the {eigenvector}. {Here we emphasize that the eigenvectors here are not standard predetermined functions but depend on the specific trajectory of the measurements.}

The final solution is a weighted sum of the eigenvectors of different order
\begin{equation}
B_r^{r_d}=\sum_i \beta_iB_i^{r_d}, \quad i=1, 2, ...
\end{equation}
{ here $\beta_i$ are the weights of the eigenvector. Both $\beta_i$  and $B_i$ can be computed with the singular-value-decomposition (SVD) \citep[e.g.][also see Appendix B]{Jackson1972, Connerney1981, ASTER2013}}.

\begin{figure}[h]
\centering
%
\includegraphics[width=0.475\textwidth]{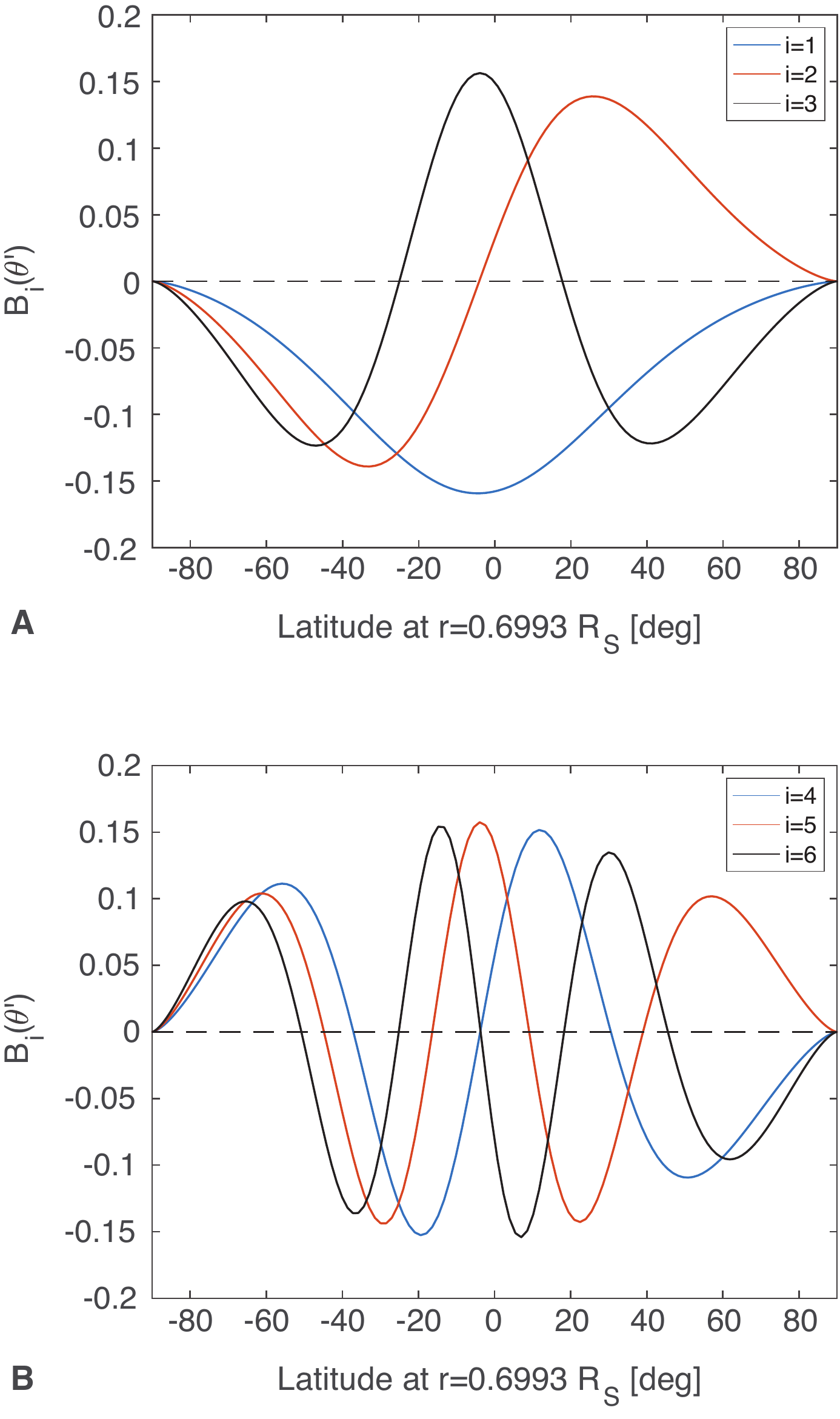}
\caption{First six {eigenvectors of the magnetic Green's function at $r=$0.6993 $R_S$ (the polar radius of the $a=0.75$ $R_S$, $c=0.6993$ $R_S$ ellipsoidal surface).} It can been seen that the eigenfunctions constructed from the Green's function feature zero values at the poles, in contrast to the $m=0$ Legendre functions which peak at the poles.}
\label{fig:eigenGreen}
\end{figure}

We choose the Gauss-Legendre quadrature points with 180 grids in the latitudinal direction to ensure high-precision integration for smooth functions. In Fig. \ref{fig:eigenGreen}, we show the first six {eigenvectors in parameter space} derived along the trajectory of the Cassini Grand Finale orbits. It can be seen that all {eigenvectors feature} zero $B_r^{r_d}$ at the poles, in contrast to the $m=0$ associated Legendre functions (the basis functions for axisymmetric Gauss coefficients) which all peak at the poles. It becomes immediately clear that with the given trajectory, the Green's function method seeks solutions with zero $B_r$ at the poles, which is an intriguing {mathematical} property of this method. Given this property and the fact that Saturn's internal magnetic field is predominantly dipolar, we employ the Green's function method to seek small-scale internal magnetic field solutions beyond spherical harmonic degree 3. 

\subsubsection{Small-scale features in Saturn's internal magnetic field from Green's function inversion} 

We adopt {the degree 1 to 3 Gauss coefficients from the Cassini 11 model} as the basis model, and seek the internal magnetic field beyond this basis model using the Green's function. To obtain a smooth solution, one needs to either truncate the solution at a certain order $i_{max}$ (see Appendix B for more details) or apply certain form of regularization. Here we choose to truncate the solution at $i_{max}$ as a first step. The truncation order of the eigenfunction, $i_{max}$, is determined by the RMS residual and the model-data misfit. 

\begin{figure*}[t]
\centering
%
\includegraphics[width=0.90\textwidth]{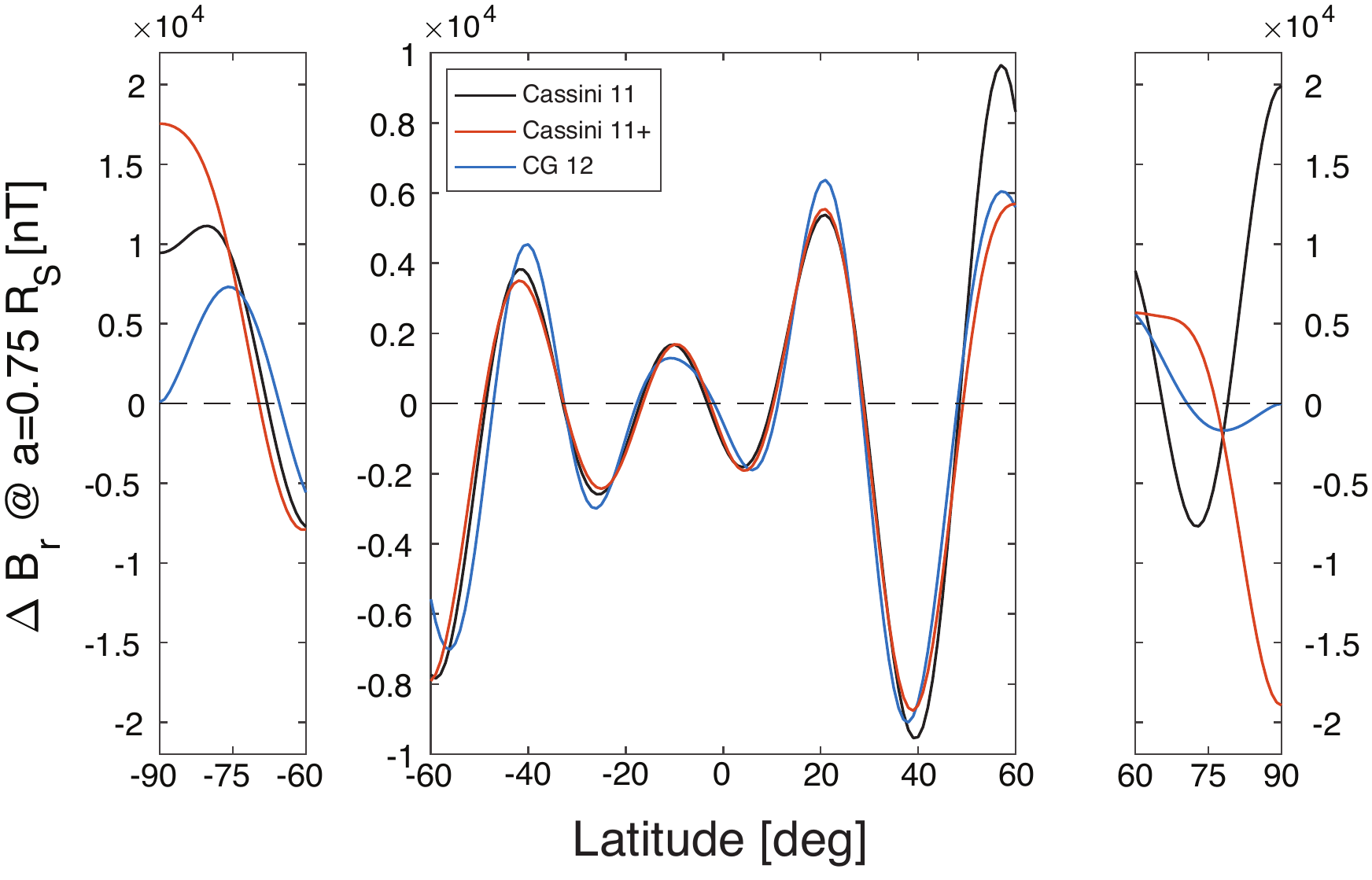}
\caption{Small-scale ($n>3$) magnetic field {of Saturn viewed at the $a=0.75$ $R_S$, $c=0.6993$ $R_S$ isobaric surface} constructed from regularized Gauss coefficients inversion (Cassini 11+ model) and from the Green's function inversion (CG 12 model).}
\label{fig:dBr_CG12_C11}
\end{figure*}

Fig. \ref{fig:dBr_CG12_C11} shows the small scale magnetic field beyond spherical harmonic degree 3, $\Delta B_r$, constructed from the Green's function with {$r_d=0.6993$ $R_S$ and $i_{max}=12$}, which we refer to as CG12 model, in which C stands for Cassini, G stands for Green's function, and 12 indicates the truncation order of the eigenfunction. This truncation order is chosen to yield a similar RMS residual to that of the Cassini 11+ model. The perturbation field from the Cassini 11+ model and the Cassini 11 model are shown in Fig. \ref{fig:dBr_CG12_C11} for comparison (the same degree-3 model has been removed for a fair comparison). It can be seen from Fig. \ref{fig:dBr_CG12_C11} that the field structures {constructed from two different methods} are very similar within $\pm$60 degrees: there are four latitudinal magnetic field bands between the equator and 60$^\circ$ latitude in each hemisphere. Above $\pm$60$^\circ$, the solution from the Green's function features zero $B_r$ at the poles (an intrinsic property of the method) while the Cassini 11$+$ model features comparable {$\Delta B_r/|B|$} to that within $\pm$ 60$^\circ$ (which results from the chosen regularization). Although the difference between the two models beyond $\pm$60$^\circ$ latitude originates from the intrinsic properties of the methods, this nonetheless highlights the non-uniqueness in the solution beyond $\pm$60$^\circ$ latitude. This non-uniqueness in the polar region should be kept in mind when interpreting the resultant $\Delta B_r$.

Once we obtain $B_r$ {at $r=r_d$}, the corresponding Gauss coefficients can be easily computed via a surface integration given the orthogonality of the spherical harmonics on a sphere. 
\begin{equation}
g_n^0=\frac{2n+1}{2(n+1)}\left(\frac{r_d}{R_P}\right)^{n+2}\int_0^\pi B_r P_n^0(\cos \theta)\sin\theta d\theta,
\label{eqn:Br2gn}
\end{equation}
where the pre-factor results from the Schmidt-normalization. Supplementary Table 1 compares the Gauss coefficients of the Green's function model (the CG12 model) to that of the Cassini 11 model \citep[]{DC2018} and the Cassini 11$+$ model. For the CG12 model, the degree 1-3 Gauss coefficients are the sum of the basis model and those computed from Eq. (\ref{eqn:Br2gn}). It can be seen that the Gauss coefficients of these models are also broadly similar: beyond degree 3, all models feature a strong and positive $g_4^0$ and a strong and negative $g_7^0$. 

\section{Electromagnetic induction response from Saturn's interior}

Electromagnetic (EM) induction can be employed to probe the interiors of planetary bodies. Examples of planetary applications of this technique include the discovery of the subsurface ocean inside Europa and Callisto from Galileo magnetometer measurements \citep[]{Khurana1998}, constraints on lunar core size from Apollo 12 and Explorer 35 magnetometer measurements \citep[]{Hood1982}, and constraints on water content variations in the mantle transition zone inside the Earth \citep[]{Kelbert2009}.

\begin{figure}[htbp]
\centering
\includegraphics[width=0.475\textwidth]{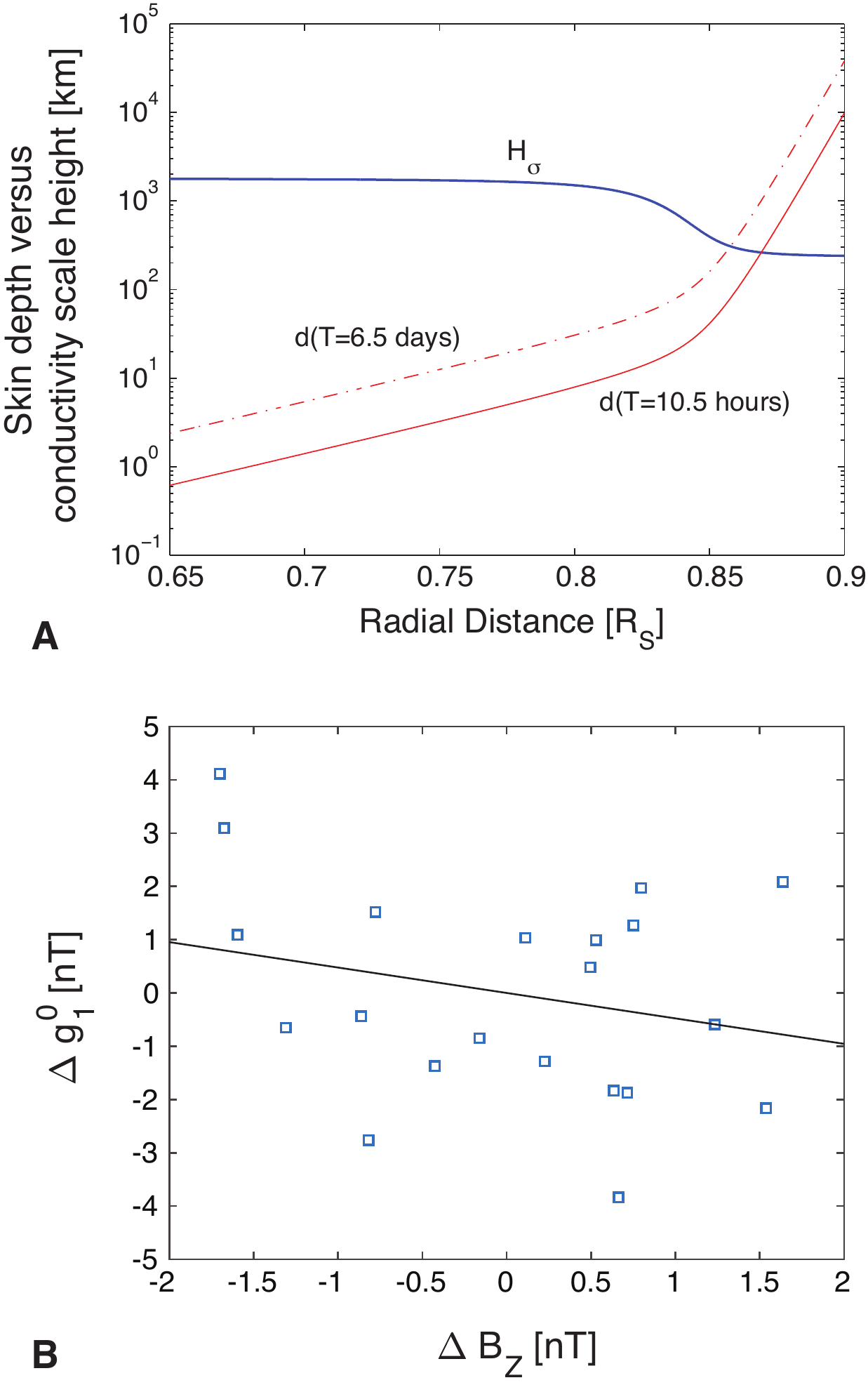}
\caption{Electromagnetic induction response from the interior of Saturn. Panel A shows the skin depth versus the electrical conductivity scale-height. It can be seen that for inducing field with frequencies between the spin frequency of Saturn and the orbital frequency of the Cassini Grand Finale orbits, the skin depth becomes comparable to or smaller than the local conductivity scale height around 0.86 $R_S$. Panel B shows the {orbit-to-orbit varying internal dipole $\Delta g_1^0$ as a function of the orbit-to-orbit varying magnetodisk field $\Delta B_Z$} derived from the Cassini Grand Finale MAG measurements. The expected induction response from an induction depth at 0.86 $R_S$ is overplotted.}
\label{fig:g10_ind_Bz}
\end{figure}

The key parameter in the EM induction is the skin-depth, $d=\sqrt{2/\omega_{ind}\mu_0\sigma}$, which depends on the frequency of the inducing field $\omega_{ind}$ and the local electrical conductivity $\sigma$. $\mu_0$ is the magnetic permeability. Since the electrical conductivity is expected to rise continuously yet rapidly as a function of depth inside Saturn \citep[]{Weir1996, Liu2008, CS2017Icarus}, the EM induction response is expected to occur at different depths for inducing fields with different frequencies. The depth at which the EM induction occurs is where the frequency dependent skin-depth $d_{ind}$ becomes comparable to or smaller than the local scale-height of the electrical conductivity $H_\sigma=\left| \sigma/\frac{d\sigma}{dr}\right|$. Given our current understanding of the electrical conductivity profile inside Saturn based on a band-closure model \citep[]{Liu2008}, EM induction is expected to occur at $r_{ind}$ around $0.87 R_S$ and $0.86 R_S$ for sounding frequencies equal to the rotational frequency of Saturn ($\sim$ 10.5 $hr$) and the orbital frequency of Cassini Grand Finale orbits (6.5 $Earth$ $days$) respectively (Fig. \ref{fig:g10_ind_Bz}A). The electrical conductivity at these depths are about 0.1 $S/m$ and 1 $S/m$ respectively. The depth from the 1-bar atmosphere is about 8000 $km$.

The magnetodisk $B_Z$ field (Table 3) is expect to induce an internal axial dipole $g_1^0(ind)$ inside Saturn. This induction response consists of two parts, a time-stationary part and a time-varying part. The magnetodisk field has a well defined mean component of order 10 $nT$, which seems to be stable over at least decadal time-scales with available in-situ observations. Given the very high electrical conductivity in Saturn's deep interior, an induction response to the stable part of the magnetodisk $B_Z$ field is expected. However, this induction response cannot be effectively separated from a stable internal axial dipole.

Thus, in searching for an induction response from the interior of Saturn, we focus on the expected time-varying part. The expected time-varying induction response $\Delta g_1^0$ to the time-varying part of the magnetodisk field $\Delta B_Z$ is that 
\begin{equation}
\Delta g_1^0 = -\frac{1}{2} \left( \frac{r_{ind}}{R_S} \right)^{1/3} \Delta B_Z.
\label{eqn:g10_Bz}
\end{equation} 
This corresponds to an induction response in which the induced radial field $B_r^{ind}$ perfectly cancels the radial component of the external inducing field $B_r^{ext}$ at $r_{ind}$. Note that the induced tangential component {$B_\theta^{ind}$} acts to {increase the external tangential component by 50\%} instead of canceling it at $r_{ind}$. The factor $1/2$ in Eq. \ref{eqn:g10_Bz} originates from the normalization of the associated Legendre polynomials which is part of the definition of $g_1^0$. Thus, the slope of $\Delta g_1^0$ versus $\Delta B_Z$ reveals the depth at which the induction response occurs. For an induction depth at $0.87 R_S$ ($0.86 R_S$), the expected slope is $-0.4773$ ($-0.4755$). 

We solve for $\Delta g_1^0$ orbit by orbit after removing the Cassini 11$+$ model and the magnetodisk field. Figure \ref{fig:g10_ind_Bz}B shows $\Delta g_1^0$ as a function of the time-varying magnetodisk $\Delta B_Z$ field orbit-by-orbit. { With the available data an induction signal seems present. If one performs a formal inversion analysis on this dataset, the expected slope is within 1$\sigma$ of that from the formal inversion analysis. However, the large scatter in the data precludes any definitive constraint on the induction depth.}

\section{Orbit-to-orbit variations in Saturn's ``internal" quadrupole magnetic moments}

\begin{figure}[htbp]
\centering
%
\includegraphics[width=0.475\textwidth]{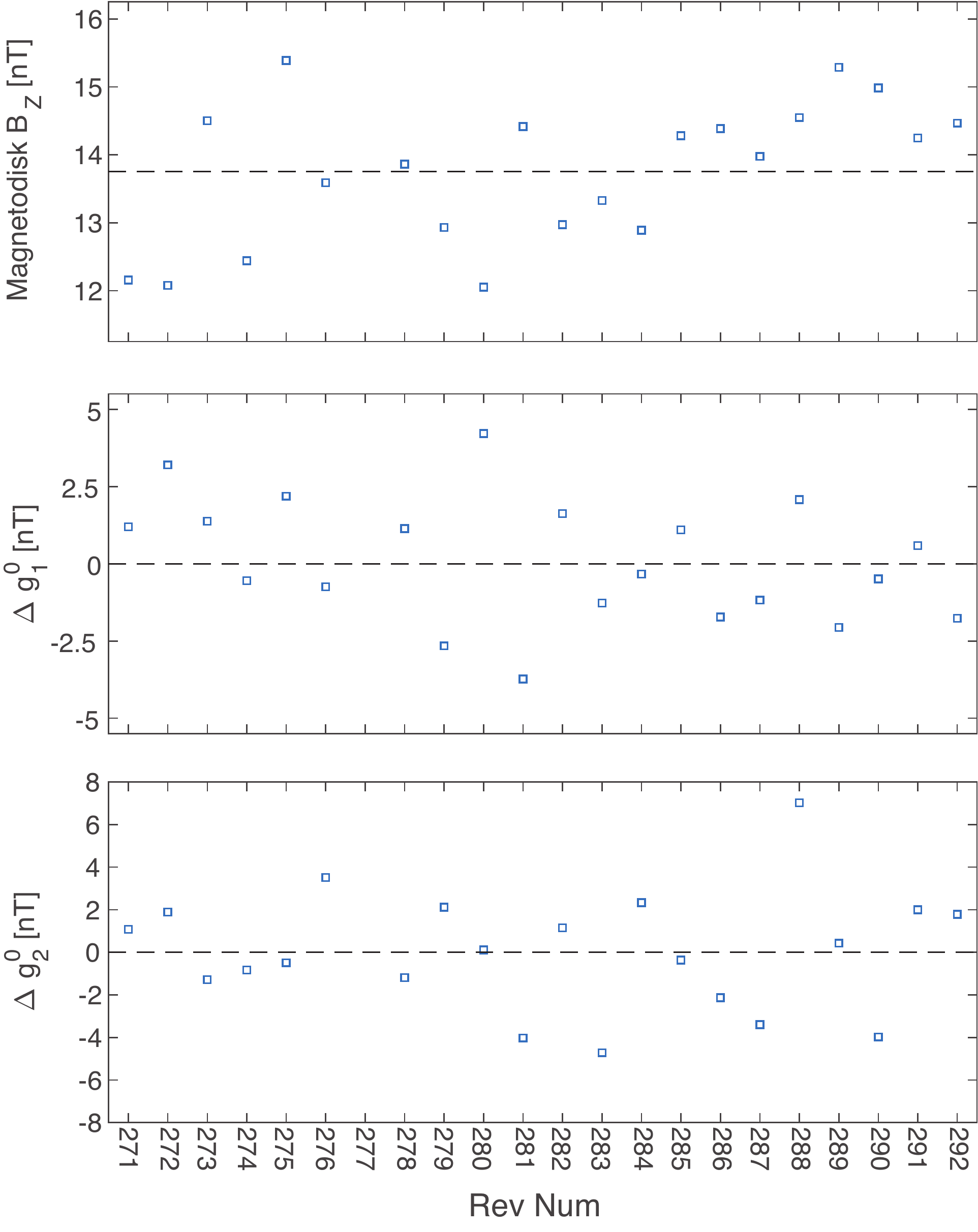}
\caption{Orbit-to-orbit variations in Saturn's external magnetodisk field, ``internal" dipole, and ``internal" quadrupole coefficients.}
\label{fig:Delta_g10_g20_Rev}
\end{figure}

In addition to solving for $\Delta g_1^0$ orbit by orbit, we also attempted to solve for $\Delta g_2^0$ orbit by orbit and found some non-negligible variations. Solving for $\Delta g_2^0$ does improve the data-model misfits, while solving for $\Delta g_n^0$ with $n>2$ does not reduce the data-model misfit much further. We attempted to solve for $\Delta g_1^0$ and $\Delta g_2^0$ separately and simultaneously, and observed negligible differences in the {resulting} values. Table 6 lists the resultant $\Delta g_2^0$, which are also plotted against Rev Number in Fig. \ref{fig:Delta_g10_g20_Rev}. It can be seen that the variations in $g_2^0$ stay within $\pm$ 4.6 $nT$, except along Rev 288 where a factor of 1.5 larger variation in $g_2^0$ were observed. Near simultaneous Hubble Space Telescope (HST) observations of the northern far-ultraviolet aurorae of Saturn recorded a strong intensification of total auroral power in the $H_2$ bands close to the periapsis time of Rev 288 \citep[]{Lamy2018}. 

\begin{table}
\caption{{Orbit-to-orbit varying Internal Dipole and Quadrupole Coefficients Measured along the Cassini Grand Finale Orbits}}
\centering
\begin{tabular}{c c c}
\hline
 Rev Num & $\Delta g_1^0 [nT]$ & $\Delta g_2^0 [nT]$\\
\hline
  271  & 1.2	& 1.1		\\ 
  272  & 3.2  	& 1.9		\\
  273  & 1.4  	& -1.3	\\
  274  & -0.5  	& -0.8	\\
  275  &  2.2  	& -0.5	\\
  276  & -0.7	& 3.5 	\\
  278  & 1.1 	& -1.2	\\
  279  & -2.7  	& 2.1 	\\
  280  & 4.2 	& 0.1		\\
  281  & -3.7 	& -4.0	\\
  282  & 1.6	& 1.1		\\
  283  & -1.3 	&  -4.6 	\\
  284  & -0.3  	& 2.3  	\\
  285  & 1.1	& -0.4 	\\
  286  & -1.7   	&  -2.1   	\\
  287  & -1.2   	& -3.4  	\\
  288  & 2.1  	& 7.0		\\
  289  & -2.1  	& 0.4		\\
  290  & -0.5 	& -4.0	\\
  291  &  0.6 	& 2.0		\\
  292  & -1.8 	& 1.8		\\
 \hline
\end{tabular}
\end{table}

Moreover, $\Delta g_1^0$ and $\Delta g_2^0$ do not exhibit strong correlation: the coefficients of correlation between the two is only {34\%}. The variability in $\Delta g_2^0$ is larger than that in $\Delta g_1^0$. The standard deviation of $\Delta g_2^0$ is {2.8 $nT$ (2.4 $nT$} if Rev 288 is excluded), while the standard deviation of $\Delta g_1^0$ is 2.0 $nT$. We speculate that the observed variations in $g_2^0$ mostly reflect variations in the east-west (zonal) currents in the ionosphere. The quadrupole moment $g_2^0$ corresponds to north-south antisymmetric zonal currents: e.g. a positive $g_2^0$ is consistent with eastward current in the north and westward current in the south. The order 5 $nT$ amplitude is consistent with our order-of-magnitude estimations of the ionospheric Hall current contributions (see Appendix C), while the pattern indicates stronger north-south asymmetry compared to the expectation of continuing the 1-bar wind pattern up to the 1100 $km$ altitude ionospheric layer. 

\section{Search for non-axisymmetry in Saturn's internal magnetic field}

As demonstrated in the analysis of Saturn's magnetic equator positions (section 3), the level of departure from perfect axisymmetry is likely only on the order of $3\times10^{-4}$. Nonetheless, we performed a search for the non-axisymmetric internal magnetic moments of Saturn based on the Cassini Grand Finale MAG measurements. The traditional Gauss coefficients representation is adopted, and the maximum SH degree and order for the non-axisymmetric moments are both set to be 3. Since the deep interior rotation rate of Saturn remains uncertain {\citep[]{Anderson2007,Read2009,Mankovich2019,Militzer2019}}, we surveyed a wide range of possible rotation periods from 10h30m00s to 10h55m00s. 

\begin{figure}[htbp]
\centering
%
\includegraphics[width=0.475\textwidth]{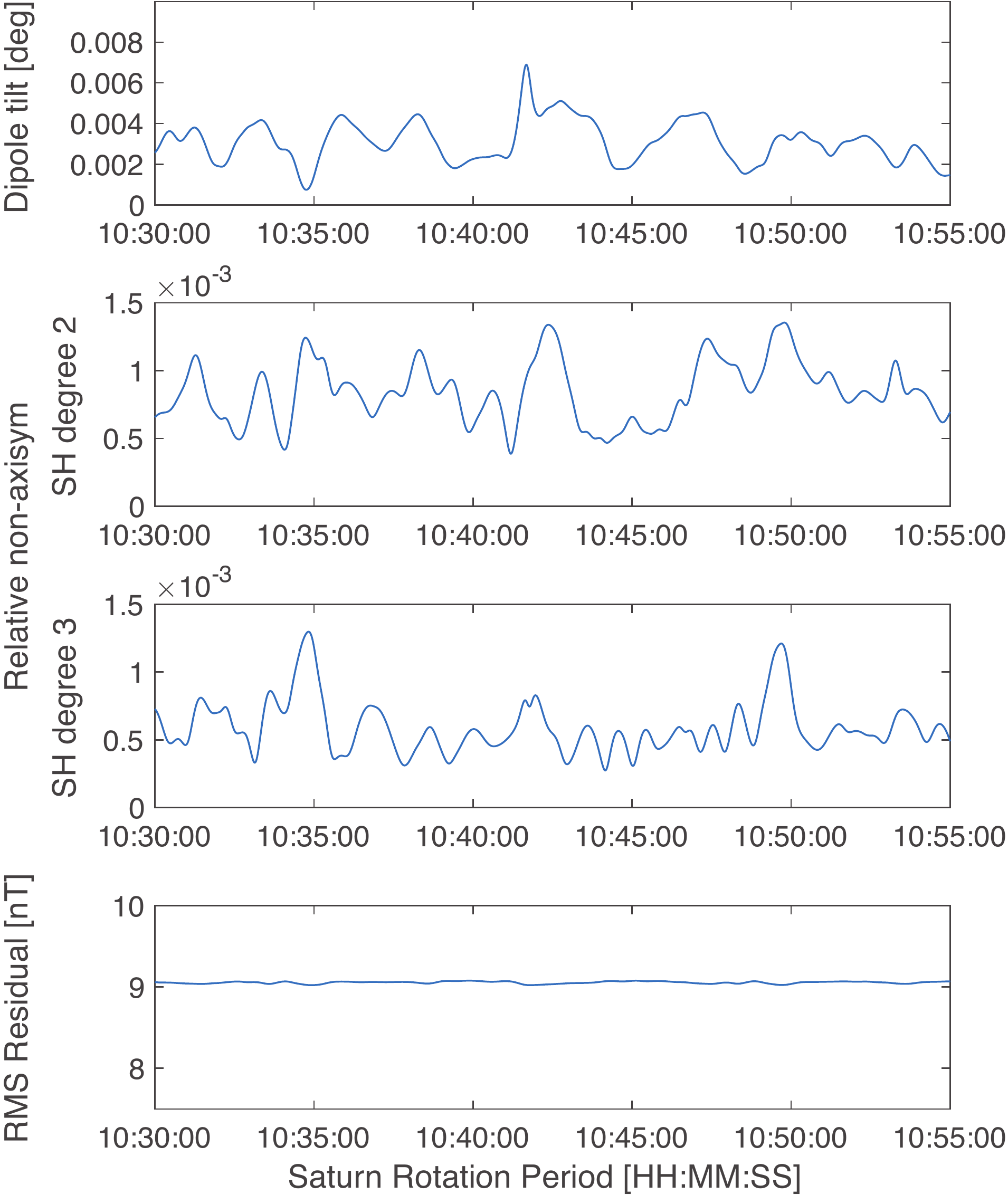}
\caption{Results from the search for non-axisymmetry in Saturn's internal magnetic field based on the Cassini Grand Finale MAG measurements. Panel A shows the dipole tilt, panel B and C show the relative non-axisymmetry in degree 2 and degree 3 moments respectively, and Panel D shows the RMS residual. All quantities are shown as a function of the assumed rotation period of Saturn's deep interior. No dominant peak in internal non-axisymmetry can be identified, and the peak dipole tilt is less than 0.007$^\circ$ {(25.2 arcsecs)}.}
\label{fig:N123_SatRot}
\end{figure}

Fig. \ref{fig:N123_SatRot} shows the dipole tilt, the relative non-axisymmetry in SH degree 2 and 3 (defined as the ratio of the amplitude of the non-axisymmetric magnetic moments to that of the axisymmetric magnetic moment of the same degree), and the RMS residual from the search. No dominant peak in the amplitude of the internal non-axisymmetric can be identified, and the peak dipole tilt is less than 0.007 degrees {(25.2 arcsecs)}. The relative non-axisymmetry in degree 2 and 3 are less than $1.5 \times 10^{-3}$. Thus, Saturn's internal magnetic field is 1000 times more axisymmetric compared to those of Earth and Jupiter. What makes Saturn's internal magnetic field so drastically different? We discuss this in the next section. 

\section{Implication for Saturn's interior}

\subsection{Magnetic axisymmetry and deep stable stratification inside Saturn}

The exceptional level of axisymmetry in Saturn's internal magnetic field revealed by the Cassini Grand Finale MAG measurements presents a challenge and an opportunity. The challenge is to our understanding of natural dynamos while the opportunity is to decode Saturn's interior structure and dynamics. Cowling's theorem {\citep[]{Cowling1933, BC1956, Hide1982}} precludes a perfectly axisymmetric magnetic field to be maintained by natural dynamos, although no lower bound on the departure from axisymmetry has been placed by this theorem. Furthermore, Cowling's theorem is a statement about the entire magnetic field in the dynamo region, much of which we cannot observe {(e.g., the toroidal field)}. Setting Cowling's theorem aside for now, Saturn's axisymmetric internal magnetic field appears special from {the perspectives of} both observations and modern understanding of the planetary dynamo process. 

From observations, highly axisymmetric magnetic fields are rare among planets. Both Earth and Jupiter feature $\sim 10^\circ$ dipole tilt, while Uranus and Neptune feature $\sim 50^\circ$ dipole tilt and strong non-axisymmetric quadrupole and octopole fields. The case of Mercury and Ganymede are less clear at this stage. Mercury's magnetic equator positions do feature $\sim$ 100 $km$ peak-to-peak variations \citep[see Fig. 4 in ][]{Anderson2012}, which are much bigger variations compared to that of Saturn given the relative small size of Mercury ($R_{Mercury}=2439.7 km$). However, whether such variations are due to  {internal non-axisymmetry or magnetospheric processes \citep[]{Jia2015}} remains to be clarified. The ESA-JAXA BepiColombo mission is expected to help resolve this issue. The non-axisymmetry of Ganymede's internal magnetic field is less clear due to the ambiguity in separation of the dynamo-generated internal field and the EM induced field {given} the limited spatial-temporal coverage of Galileo {Ganymede flybys} \citep[]{KIVELSON2002}. The ESA JUpiter ICy moons Explorer (JUICE) mission is expected to resolve this ambiguity {with low-altitude Ganymede orbits}. 

From modern understanding of the planetary dynamo process, highly axisymmetric magnetic fields are rare in convective dynamo simulations. Highly supercritical rotating convection is strongly non-axisymmetric. Due to inverse cascade \citep[]{Guervilly2014, Rubio2014}, the non-axisymmetry in the convective flows tends to have strong large-scale components. These large-scale non-axisymmetric convective flows are expected to generate large-scale non-axisymmetric magnetic fields as observed in the majority of convective numerical dynamo simulations. {In numerical dynamo surveys,} the magnetic field {in} the dipolar branch tends to feature a modest amount of non-axisymmetry, e.g. with dipole tilt between 5 to 10 degrees, while the magnetic field {in} the multi-polar branch tends to be dominated by non-axisymmetry \citep[]{Christensen2006, Soderlund2012, Duarte2013}. 

The most appealing mechanism to axisymmetrize Saturn's internal magnetic field is via the combination of strong differential rotation and suppression of large-scale non-axisymmetric convective motion on top of the dynamo region \citep[]{Stevenson1980, Stevenson1982}. It should be emphasized that the differential rotation here refers to the shear between the flow in the convective dynamo region and the flow {in an electrically conducting layer above the convective dynamo region}. In principle, only differential rotation in the spherical radial direction is needed. Such differential rotation tends to destroy non-axisymmetric magnetic features via advectively shearing them, then diffusively dissolving them. Under the case of angular velocity as a function of radial distance only and ignoring the dynamic feedback from the Lorentz force induced, this process can be thought of as electromagnetic filtering. In addition to strong differential rotation on top of the deep dynamo, suppression of large-scale non-axisymmetric convective motion outside the deep dynamo is a necessary ingredient to maintain an axisymmetric magnetic field, since any large-scale non-axisymmetric convective motion {in an electrically conducting region} would lead to large-scale non-axisymmetric magnetic field. The most likely way these two conditions are satisfied inside Saturn is via the formation of a stably stratified \citep[]{Stevenson1980} or double diffusively convecting \citep[]{LC2012, LC2013} layer on top of the deep fully convective dynamo. Helium rain {\citep[]{Stevenson1975,Stevenson1977, Morales2009, Lorenzen2009}} could lead to the formation of such a layer. However, the picture of helium rain inside Saturn is in doubt since we lack a direct measurement of significant helium depletion in the atmosphere {of Saturn}. The established helium depletion in Jupiter from Galileo results and the expected lower entropy in Saturn suggests helium rain should occur in Saturn to a greater extent than in Jupiter but this is contingent on the standard assumption of isentropy down to the {pressure} level of helium insolubility in both planets. Other processes inside Saturn could lead to the formation of such a layer on top of the dynamo. For example, if dissolved core material (heavy elements) is convectively mixed upward to around 0.6 $R_S$, this would create a stable compositional gradient near this depth since the layer above would feature less heavy elements. The thickness of this layer and the format of radial motion in this layer, e.g. oscillatory motion or small-scale double diffusive convective motion, is determined by the competition between the thermal gradient and the compositional gradient \citep[]{LC2012}. The measured extreme level of axisymmetry in Saturn's magnetic field can help us constrain these properties. We loosely refer to this layer as a ``stable layer" even though it should be understood that this layer could be double diffusively convecting. 

\begin{figure}[htbp]
\centering
%
\includegraphics[width=0.475\textwidth]{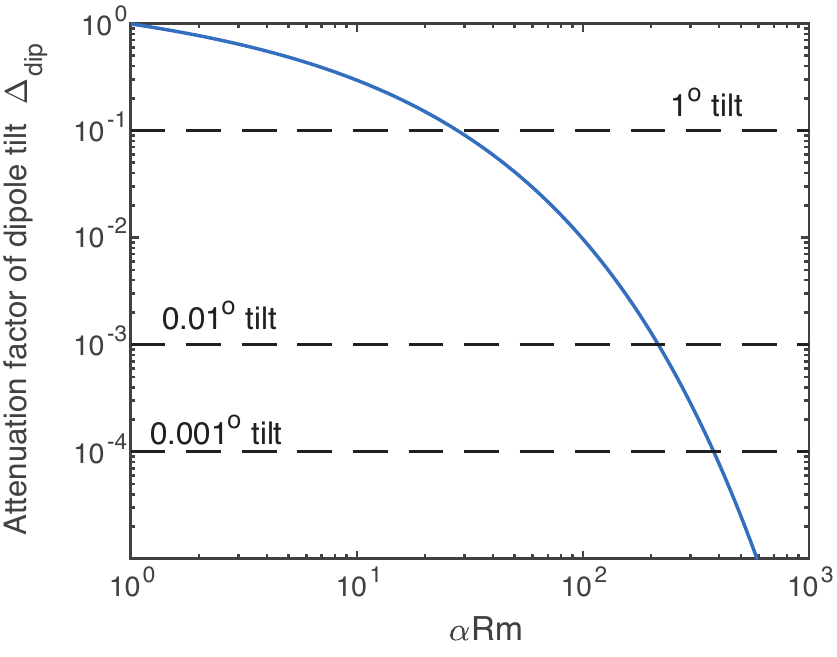}
\caption{The attenuation factor of the internal dipole tilt as a function of $\alpha Rm$ according to the kinematic plane-layer model by \citet[]{Stevenson1982}. To reach a 0.007$^\circ$ dipole tilt, $\alpha Rm$ needs to be larger than 238. The stable layer needs to be thicker than 2500 $km$ (5600 $km$) if the differential rotation between the deep dynamo and the stable layer is about 5 $mm/s$ (1 $mm/s$).}
\label{fig:alpha_Rm_Saturn}
\end{figure}

An important non-dimensional parameter to quantify the stable layer's ability to axisymmetrize the dynamo generated magnetic field is 
\begin{equation}
\alpha Rm=\frac{mL_{Stable}}{R_{Dynamo}}\frac{\Delta u_\phi L_{Stable}}{\eta_{Stable}},
\end{equation}
here $m$ is the azimuthal wave number (spherical harmonic order $m$), $L_{Stable}$ is the thickness of the stable layer, $R_{Dynamo}$ is the radius of the deep dynamo, $\Delta u_\phi$ is the differential rotation between the stable layer and the deep dynamo, and $\eta_{Stable}$ is the magnetic diffusivity of the stable layer. Fig. \ref{fig:alpha_Rm_Saturn} shows the maximum attenuation factor of the dipole tilt ($m=1$), {which is the ratio of the dipole tilt above the stable layer to that below the stable layer}, as a function of $\alpha Rm$ according to the plane layer kinematic model of \citet[]{Stevenson1982}: 
\begin{equation}
\Delta_{max}=\frac{1.59}{(\alpha Rm)^{1/12}}exp\left[-\sqrt{2}/3 \left(\alpha Rm\right)^{1/2} \right].
\end{equation}

Assuming a 10$^\circ$ dipole tilt in the deep dynamo region, to achieve the observed upper limit of dipole tilt, 0.007$^\circ$, outside the stable layer, $\alpha Rm$ needs to be larger than 238. If we assume 1 $mm/s$ (5 $mm/s$) differential rotation between the stable layer and the deep dynamo and a magnetic diffusivity of 4 $m^2/s$ (equivalent to an electrical conductivity of $2\times10^5$ $S/m$) and a deep dynamo radius around 0.55 $R_S$, this requires a stable layer thicker than 5600 $km$ (2500 $km$). It should be immediately realized that a ``stable" layer over 2500 $km$ thick cannot be a purely diffusive layer. Assuming a thermal conductivity of 100 $W/K/m$ \citep[]{French2012}, to diffusively transport the observed luminosity 2 $W/m^2$ of Saturn through a purely conducting layer over 2500 $km$ thick around 0.55 $R_S$ would require a thermal gradient as large as {66 $K/km$} or a temperature jump over {165000 $K$} across the stable layer. Thus, double diffusive convection {and/or fluid waves} must be present to transport the heat out.

Moreover, $\alpha Rm$ and the ``stable" layer thickness derived here is likely a lower limit. In this kinematic model \citep[]{Stevenson1982}, the dynamical feedback from the magnetic field to the flow {via the Lorentz force} was ignored. {Such dynamical feedback} likely would reduce the efficiency of axisymmetrization. Whether a very large $\alpha Rm$ can be achieved in a fully dynamic situation is unclear, since the differential rotation between the stable layer and the deep dynamo $\Delta u_\phi$ {would be} dynamically constrained. In published Saturn dynamo simulations with a stable layer \citep[]{CW2008, Stanley2010}, $\alpha Rm$ is on the order of 15 or less, consistent with the $\sim 1^\circ$ dipole tilt achieved. Whether there is a dynamical limit on $\alpha Rm$ and the axisymmetrization efficiency of this mechanism remains an open question for future investigations. 

\subsection{Banded magnetic perturbations and deep zonal flows in the semi-conducting layer of Saturn}

\begin{figure*}[htb]
\centering
%
\includegraphics[width=0.8\textwidth]{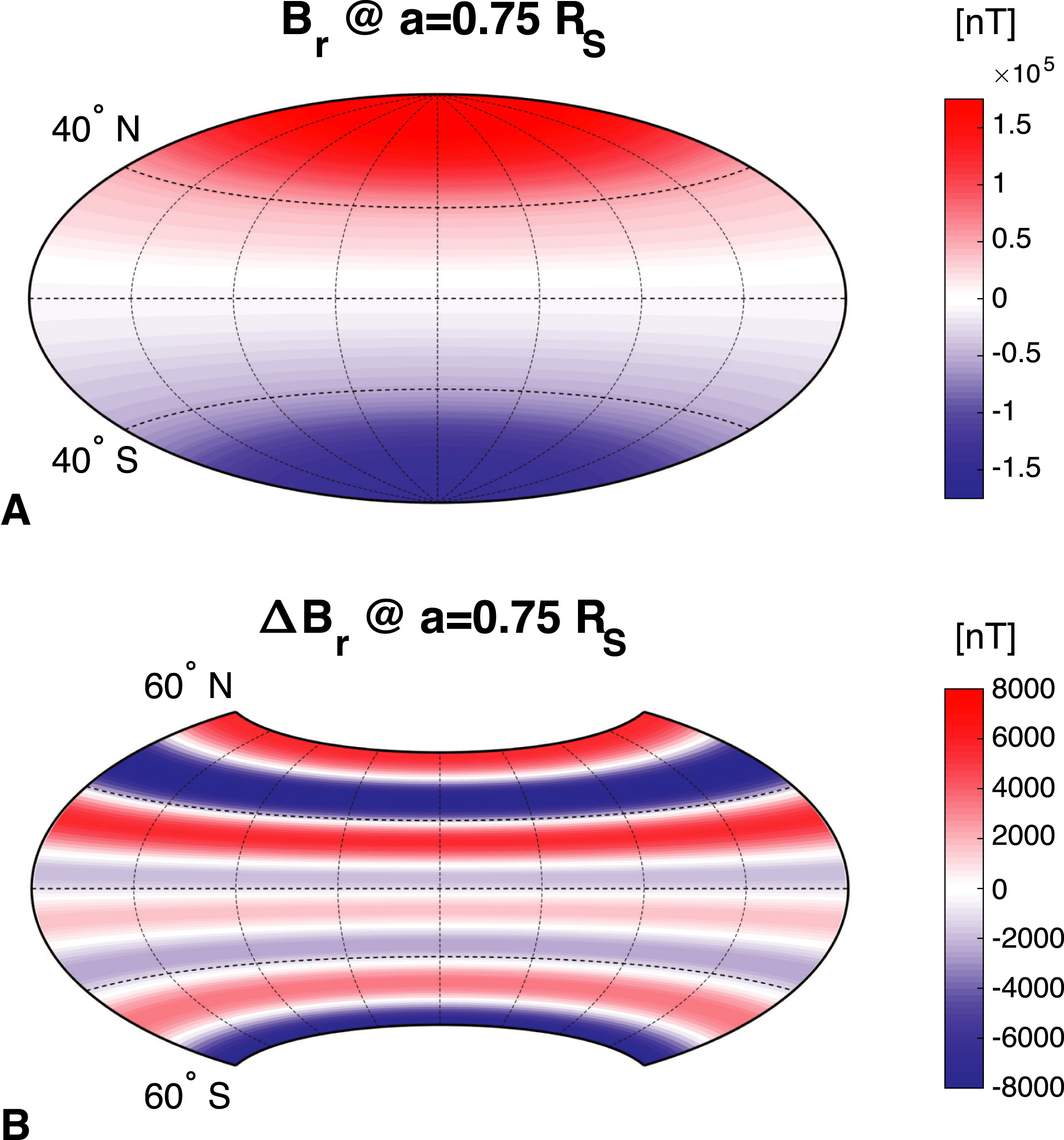}
\caption{{Saturn's large and small scale radial magnetic field at the $a=0.75$, $c=0.6993$ $R_S$ isobaric surface according to the Cassini 11+ model. Saturn's large scale radial magnetic field at this depth features a relatively weak equatorial region, $B_r$ remains less than 50,000 $nT$ ($<$1/3 of its peak value) between $\pm$40$^\circ$. Saturn's small-scale magnetic field at this depth features eight alternating bands between $\pm$60$^\circ$, with typical amplitude of $\sim$ 5\% - 10\% of the background field.}}
\label{fig:Br_dBr_C11p_075a}
\end{figure*}

It is intriguing that although Saturn's internal magnetic field appears to be perfectly axisymmetric, it does feature a rich axisymmetric magnetic spectrum extending to spherical harmonic degree 9 and beyond. The degrees 1 to 3 magnetic moments likely originate from the deep dynamo { given their order-of-magnitude power dominance over that of the higher degree moments when viewed at 0.75 $R_S$}. The magnetic moments beyond degree 3 {and the associated latitudinally banded magnetic perturbations likely originate from a shallow secondary dynamo with alternating bands of deep zonal flows in the semi-conducting layer of Saturn. As shown in \citet[]{CS2017Icarus}, banded differential rotation and local helical motion in the semi-conducting region could generate a rich axisymmetric magnetic spectrum even if the deep dynamo field is simply an axial dipole. The Cassini MAG data suggests that there are eight alternating bands of magnetic perturbations between $\pm$ 60$^\circ$ at the $a=0.75 R_S$ elliptical surface (Fig. \ref{fig:dBr_CG12_C11} \& \ref{fig:Br_dBr_C11p_075a}B). The typical latitudinal width of each magnetic band is $\sim$ 15$^\circ$. If we project the observed 1-bar surface zonal winds along the direction of the spin-axis towards the $a=0.75 R_S$ elliptical surface, there are eight alternating bands of zonal jets between $\pm$ 60$^\circ$ with the off-equatorial jets feature typical latitudinal width $\sim$ 15$^\circ$ at this depth. Thus, the characteristic width of the latitudinally banded magnetic perturbations is similar to that of the $Z$-projection of the surface off-equatorial zonal jets.

Three necessary ingredients for a secondary dynamo in the semi-conducting layer are 1) the existence of a deep dynamo which provides the background magnetic field $\mathbf{B_0}$, 2) differential rotation in the semi-conducting layer which produces toroidal magnetic field $\mathbf{B_T}$ from $\mathbf{B_0}$ through the dynamo $\omega$-effect, and 3) local helical motion which produces observable poloidal magnetic field perturbations $\Delta\mathbf{B_P}$ from $\mathbf{B_T}$ through the dynamo $\alpha$-effect \citep[]{Parker1955, SKR1966, SK1966}. Heat transport requirements and background rotation naturally lead to helical motion and local dynamo $\alpha$-effect in the semi-conducting layer. The spatial profile of the resultant $\mathbf{B_T}$ and $\Delta\mathbf{B_P}$ are expected to be spatially correlated with that of the differential rotation. The fact that the characteristic width of the latitudinally banded magnetic perturbations is similar to that of the $Z$-projected surface zonal jets lends further support to the idea that the profile of deep zonal flows in Saturn's semi-conducting layer strongly resemble that of the observed surface zonal jets \citep[]{Iess2019, Galanti2019, Militzer2019}. In addition to the idealized mean-field model \citep[]{CS2017Icarus}, secondary dynamo action has also been observed in some global numerical dynamo simulations for giant planets featuring a radially varying electrical conductivity and deep zonal flows in the outer layers \citep[e.g.][]{Gastine2014,Duarte2018}.

The peak toroidal magnetic field production could occur anywhere between the top of the semi-conducting layer (e.g. $\sim$ 0.87$R_S$ where $\sigma \sim$ 0.1 $S/m$) and the base of the semiconducting layer (to be defined later), since it is determined by the competition between the decaying wind velocity and the increasing electrical conductivity as a function of depth.  Regardless of the peak production depth, the toroidal magnetic field will diffuse downward to the base of the semi-conducting layer \citep[e.g., see Figs. 2 \& 10 in][]{CS2017Icarus}. The poloidal magnetic field perturbations $\Delta\mathbf{B_P}$, however, are expected to be generated mainly near the base of the semi-conducting layer, due to its dependence on $\sigma^2$.} The ``base of the semi-conducting layer" is defined by either 1) the transition to the main dynamo, which likely occurs before the saturation of the electrical conductivity, or 2) the upper end of the ``stable layer" which provides a well-defined separation of the shallow dynamo from the deep dynamo. 

Since the secondary dynamo lies above the ``stable layer", will it generate secondary non-axisymmetric magnetic field that violate the observational constraints? The answer to this question is two-fold. First, in the spirit of mean field electrodynamics, the $\alpha$-effect is not dependent on longitude and hence does not introduce large scale non-axisymmetric field, though at the scale of the convective eddies it necessarily involves motions and small scale fields that have longitudinal dependence. However, the longitudinal dependent fields are expected to be much smaller than the axisymmetric field arising from the $\alpha$-effect. Second, a 5\% non-axisymmetry associated with the high-degree ($n>3$) magnetic moments will produce peak non-axisymmetric magnetic fields on the order of 5 $nT$ along the S/C trajectory. This likely is still compatible with the Cassini MAG measurements. 

As discussed in \citet[]{DC2018} and in \citet[]{CS2017Icarus}, the separation of the magnetic field of shallow origin from that of deep origin is not clear-cut. Taking a step-back to examine the large-scale field which most likely originates from the deep dynamo field, the fact that $g_1^0$ and $g_3^0$ take the same sign implies that the radial magnetic flux is expelled from the equatorial region and pushed towards mid-to-high latitude {(see Fig. \ref{fig:Br_dBr_C11p_075a}A)}. This could originate from a deep ``equatorial" jet either in the stable layer or in the deep dynamo region itself, which would tend to clear-out the radial flux so that the steady-state magnetic field approaches that of a Ferraro-corotation state: $\mathbf{B}\cdot\nabla\omega=0$, here $\omega$ is the local angular velocity. Also as discussed in \citet[]{DC2018}, if a significant part of the magnetic field with $n \le 9$ has a deep origin, the poles deep inside the planet (e.g. at 0.5 $R_S$) could feature almost zero radial magnetic field. Almost zero radial magnetic field at the poles at the deep dynamo surface could originate from flux expulsion and/or time-varying process inside a tangent cylinder \citep[]{SJ2005, LAO2017, Schaeffer2017, Cao2018PNAS} defined by a central core (mostly likely a stably stratified fluid core instead of a solid core inside Saturn), which does not participate in the large-scale convection in the deep dynamo. 

\section{Summary and Outlook}

We have analyzed the full Cassini Grand Finale MAG dataset with the goal to characterize and understand the internal magnetic field and interior of Saturn. Saturn's internal magnetic field turns out to be axisymmetric with respect to the spin-axis to an exceptional level; the dipole tilt which is a good proxy for the large-scale non-axisymmetry, must be smaller than 0.007$^\circ$ { (25.2 arcsecs)}. This extreme level of axisymmetry sets key constraints on the form of convection in the highly conducting layer of Saturn. A stably stratified {electrically conducting} layer thicker than 2500 $km$ above Saturn's deep dynamo {could axisymmetrize Saturn's internal magnetic field to the observed level, if the dynamical feedback from the magnetic field does not enter the leading order force/vorticity balance}. Furthermore, a heat transport mechanism other than pure conduction, e.g. double diffusive convection or waves, must exist within this layer {to be compatible with the observed luminosity of Saturn}.  

Although almost perfectly axisymmetric, there is a modest amount of north-south asymmetry {in Saturn's internal magnetic field}, directly demonstrated by the $\sim$ 5\% northward offsets of Saturn's magnetic equator from the planetary equator. In addition to the well-resolved axisymmetric low {spherical harmonic} degree $(n \le 3)$ magnetic moments, Saturn's magnetic field features an axisymmetric yet rich magnetic energy spectrum, which corresponds to {latitudinally banded magnetic perturbations when viewed at the $a=0.75$ $R_S$, $c=0.6993$ $R_S$ isobaric surface}. Such {latitudinally} banded magnetic perturbations likely arise from a ``shallow" secondary dynamo action within the semi-conducting layer of Saturn, enabled by differential rotation, small-scale helical motion, and the background magnetic field provided by the deep dynamo. Regularized inversion with spherical harmonic solutions as basis functions as well as truncated Green's function solutions demonstrated that the small-scale axisymmetric magnetic field between $\pm$60$^\circ$ latitude at the {$a=0.75$ $R_S$ non-spherical ``dynamo surface"} can be well determined, while the details of the small-scale field above $\pm$60$^\circ$ latitude are less certain. It should be noted that the area above $\pm$60$^\circ$ latitude is less than 14\% of the surface area. To fully resolve the small-scale magnetic field of Saturn above $\pm$60$^\circ$ latitude, including both the axisymmetric field and the non-axisymmetric field, low altitude magnetic field measurements directly above the polar region are needed. This task is left to future missions to the Saturn system. 

%
%
%
%
\appendix

\section{Gauss coefficients representation of the internal planetary magnetic field}

The traditional Gauss coefficients representation of the internal planetary magnetic field outside of the source region are shown here for convenience. 
\begin{equation}
V=\sum_{n=1} \sum_{m=0}^n R_p \left( \frac{R_p}{r} \right)^{n+1} \left[ g_n^m cos m\phi + h_n^m sin m\phi \right] P_n^m \left( cos \theta \right),
\end{equation}
\begin{equation}
\mathbf{B}=-\nabla V,
\end{equation}

\begin{equation}
B_r=\sum_{n=1} \sum_{m=0}^n \left( n+1 \right) \left(\frac{R_p}{r} \right)^{n+2}\left[ g_n^m cos m\phi + h_n^m sin m\phi \right] P_n^m \left( cos \theta \right),
\end{equation}

\begin{equation}
B_\theta=-\sum_{n=1} \sum_{m=0}^n \left(\frac{R_p}{r} \right)^{n+2}\left[ g_n^m cos m\phi + h_n^m sin m\phi \right] \frac{dP_n^m \left( cos \theta \right)}{d\theta},
\end{equation}

\begin{equation}
B_\phi=\sum_{n=1} \sum_{m=0}^n \left(\frac{R_p}{r} \right)^{n+2} \frac{m}{sin\theta} \left[ g_n^m sin m\phi - h_n^m cos m\phi \right] P_n^m \left( cos \theta \right),
\end{equation}
where $R_p$ is the reference radius here taken to be the 1-bar {equatorial} radius of Saturn, $\left(g_n^m, h_n^m \right)$ are the Gauss coefficients, $n$ and $m$ are the spherical harmonic degree and order respectively, $r$ is the spherical radial distance from the center of the planet, $\theta$ and $\phi$ are the co-latitude and east longitude respectively, and $P_n^m (cos \theta)$ are the Schmidt semi-normalized associated Legendre functions. 

\section{Green's function for the internal planetary magnetic field {and the eigenvectors of the inverse problem}}
As shown in Gubbins and Roberts (1983) and Johnson and Constable (1997), the mapping between the magnetic field at {a spherical} dynamo surface to anywhere above is
\begin{equation}
B_{r,\theta,\phi}^{obs}(r,\theta,\phi)=\int_0^{2\pi}\int_0^{\pi} B^{r_D}_r(\theta',\phi')G_{r,\theta,\phi}(\mu)\sin\theta'd\theta'd\phi',
\label{eqn:MagGreen3C}
\end{equation}
{where $B_r^{r_D}$ is the radial component of the magnetic field at the $r=r_D$ spherical dynamo surface,  $B^{obs}_{r,\theta,\phi}$ are three components of the internal magnetic field measured above the dynamo surface, $\theta$ is colatitude, $\phi$ is longitude, and $\mu$ is the consine of the angle between the position vectors $\hat{r}$ and $\hat{r}'$.}

The Green's function for each component are
\begin{equation}
G_r(\mu)=\frac{b^2}{4\pi}\frac{1-b^2}{f^3},
\end{equation}
\begin{equation}
G_\theta(\mu)=-\frac{b^3}{4\pi}\frac{1+2f-b^2}{f^3T}\frac{d\mu}{d\theta},
\end{equation}
\begin{equation}
G_\phi(\mu)=-\frac{b^3}{4\pi\sin\theta'}\frac{1+2f-b^2}{f^3T}\frac{d\mu}{d\phi},
\end{equation}
and 
\begin{equation}
\mu=\hat{r}\cdot\hat{r}',
\label{eqn:mu}
\end{equation}
\begin{equation}
b=\frac{r_D}{r},
\end{equation}
\begin{equation}
f=(1-2b\mu+b^2)^{1/2},
\end{equation}
\begin{equation}
T=1+f-\mu b.
\end{equation}

{The surface integration can be discretized, the forward problem can then be expressed as 
\begin{equation}
\mathbf{data} = G \: \mathbf{model},
\end{equation} in which $\mathbf{data}$ is the three component internal magnetic field at the measurement location $B_{r,\theta,\phi}^{obs}(r,\theta,\phi)$, $\mathbf{model}$ is the profile of $B_r^{r_D}$, and $G$ is the matrix expression of the integration of the Green's functions (\ref{eqn:MagGreen3C}). It should be emphasize here that $G$ is a function of the position of the measurements only.  

The inverse problem can then be computed using the generalized inversion analysis \citep[e.g.][]{Jackson1972, Connerney1981, ASTER2013}. Here we briefly explain this analysis, aiming at clarifying the meaning of the eigenvector of parameter space here. Assuming there are $n$ number of measurements and $m$ number of parameters which means discretizing the surface integration (eq. \ref{eqn:MagGreen3C}) into $m$ points on the spherical surface $r=r_D$, $\mathbf{data}$ is a $n \times 1$ vector, $G$ is a $n \times m$ matrix, and $\mathbf{model}$ is a $m \times 1$ vector. The matrix $G$ can be factored using the singular-value-decomposition into the product 
\begin{equation}
G = U \Lambda V^T,
\end{equation}
in which $U$ is a $n \times p$ matrix, $\Lambda$ is a diagonal matrix of $p$ number of non-zero eigenvalues ($\lambda_1$,$\lambda_2$,$\lambda_3$,...,$\lambda_p$), and $V$ is a $m \times p$ matrix. Each column of the $V$ matrix, $\mathbf{V_i}$, is one \textbf{eigenvector in the parameter space}. In our formulation, each $V_i$ is a profile of $B_r^{r_D}$. The solution $\mathbf{model}$ can then be computed as a weighted sum of the different eigenvectors in the parameter space
\begin{equation}
\mathbf{model}=\sum_i \beta_i\mathbf{V_i}, \quad i=1, 2, ...
\label{eqn:modelVi}
\end{equation}
which for this particular problem can be expressed as 
\begin{equation}
B_r^{r_D}=\sum_i \beta_iB_i^{r_D}, \quad i=1, 2, ...
\end{equation}
here $\beta_i$ is a weight whose value is the $i^{th}$ element of the vector $U^T\mathbf{data}$ divided by the $i^{th}$ eigenvalue $\lambda_i$: $\beta_i=\left(U^T\mathbf{data}\right)_i/\lambda_i$. In constructing the final model solution, truncation at order $i_{max}$ here simply means truncating the summation in equation (\ref{eqn:modelVi}) at order $i_{max}$.}

\section{Ionospheric Hall currents and their associated magnetic field}

Zonal flows likely exist in the ionosphere of Saturn. The intra-D ring field-aligned current as measured along the Cassini Grand Finale orbits could arise from the ionospheric Pedersen currents driven by the zonal flows. Such zonal flows would also drive ionospheric Hall currents, which would be in the zonal ($\hat{\phi}$) direction. Modeling of the measured $B_\phi$ combined with a global ionospheric conductivity profile \citep[]{MW2006, Galand2011, MW2012} indicates that amplitude of the zonal flow at the ionospheric peak conductivity layer likely is 50\% of that at 1 bar. Taking this value, we can make an order of magnitude estimation of the zonal ionospheric Hall current as
\begin{equation}
I_\phi=\Sigma_{H}|B|u_\phi,
\end{equation}
in which $\Sigma_H$ is the height-integrated ionospheric Hall conductivity ($\sim$10 $S$ near local noon at the equator), $|B|$ is the magnetic field strength, and $u_\phi$ is the zonal velocity in the ionospheric peak conductivity layer. 

Since we aim at an order-of-magnitude estimation of the magnetic field associated with the ionospheric Hall current, we assume axisymmetry as a first step. In this first step, we further assume the ionospheric Hall conductivity takes the noon values at all local times, which should yield an upper bound on the current density and the associated magnetic fields. The axisymmetric assumption is a reasonable one as long as the zonal extent of the current is much wider than the spatial coverage of the measurements. 

One can then obtain the $(B_r, B_\theta)$ associated with the zonal Hall currents via solving a boundary value problem: treating the ionospheric Hall currents as boundary currents. The boundary conditions are 
\begin{equation}
B_{r,above}=B_{r,below},
\end{equation}
\begin{equation}
B_{\theta,above}-B_{\theta,below}=\mu_0I_\phi,
\end{equation}
here $above$ and $below$ refers to above and below the ionosphere respectively.

It can be shown that above the ionosphere, the magnetic field associated with the Hall currents can be expressed as 
\begin{equation}
B_H=-\nabla V_H,
\end{equation}
\begin{equation}
V_H=\sum R_I \left( \frac{R_I}{r} \right) ^{n+1} A_n^0 P_n^0 \left( \cos \theta \right),
\end{equation}
\begin{equation}
A_n^0=-\frac{n}{2n+1}\mu_0I_\phi^n,
\end{equation}
here $R_I$ is the radial distance of the ionospheric peak conductivity layer from the center of the planet and $I_\phi^n$ is $n$-th degree coefficients of the decomposition of $I_\phi$ onto $dP_n^0/d\theta$,
\begin{equation}
I_\phi=\sum_n I_\phi^n \frac{dP_n^0(\cos \theta)}{d\theta}.
\end{equation}

The corresponding Gauss coefficients, re-normalized with respect to the 1-bar radius, are then simply
\begin{equation}
g_n^0(Hall)=A_n^0 \left( \frac{R_I}{R_P} \right)^{n+2}.
\end{equation}

\begin{table}
\caption{Gauss Coefficients associated with zonal Hall currents in Saturn's Ionosphere}
\centering
\begin{tabular}{c c}
\hline
  [nT] &   \\
\hline
  $g_1^0(Hall)$  & 6    \\
   $g_2^0(Hall)$  & 0.06   \\
   $g_3^0(Hall)$  & -4.15    \\
   $g_4^0(Hall)$  & -0.24    \\
   $g_5^0(Hall)$  & 2.55    \\
   $g_6^0(Hall)$  & 0.22   \\
   $g_7^0(Hall)$  & -1.26 \\
   $g_8^0(Hall)$  & -0.42  \\
   $g_9^0(Hall)$  & 0.20 \\
   $g_{10}^0(Hall)$  & 0.20   \\
   \hline
\end{tabular}
\label{tab:gn0_Hall}
\end{table}

%
%

%

\section*{Acknowledgments}
We acknowledge support from the Cassini Project. Work at Imperial College London was funded by Science and Technology Facilities Council (STFC) consolidated grant ST/N000692/1. Work at the University of Leicester was funded by STFC consolidated grant ST/N000749/1. M.K.D. is funded by Royal Society Research Professorship RP140004. H.C. is funded by NASA Jet Propulsion Laboratory (JPL) contract 1579625. H.C.'s visit to Imperial College London was funded by the Royal Society grant RP 180014. E.J.B. was supported by a Royal Society Wolfson Research Merit Award. The derived model parameters are given in Tables 3 - 6 and Supplementary Table 1. We thank Burkhard Militzer for providing the interior shape of Saturn and helpful discussions. Fully calibrated Cassini magnetometer data are available at the NASA Planetary Data System at https://pds.nasa.gov.

\bibliography{Cao2019Saturn}{}
\bibliographystyle{aasjournal}

\end{document}